\documentclass[11pt,a4paper]{article}
\pdfoutput=1
\usepackage[margin=2.5cm]{geometry}
\usepackage[latin1]{inputenc}
\usepackage{amsmath}
\usepackage{amsfonts}
\usepackage{amssymb}
\usepackage{graphicx}
\usepackage{multirow}
\usepackage{cancel}
\usepackage{xcolor}
\usepackage{float}
\usepackage{wasysym}
\usepackage{physics}
\usepackage{cite}
\usepackage{comment}
\usepackage[toc,page]{appendix}
\usepackage[font=footnotesize,labelfont=bf]{caption}
\usepackage{fancyhdr}
\usepackage[colorlinks=true, linkcolor  = blue,urlcolor=blue,citecolor=violet]{hyperref}

\numberwithin{equation}{section}

\begin{document}
\vspace*{2.5cm}
\begin{center}
{\LARGE {Interpolating geometries and the stretched dS$_2$ horizon}}
\end{center}

\vskip10mm
\begin{center}
{\small{Dionysios Anninos and Eleanor Harris}}%^{\tikz\penguin[scale=0.2];}$ $^{\tikz\owl[scale=0.2];}$}}
\end{center}

\vskip 5mm
\begin{center}
{\footnotesize{Department of Mathematics, King's College London, the Strand, London WC2R 2LS, U.K.}}
\end{center}
\vskip 5mm
\begin{center}
{\footnotesize{dionysios.anninos@kcl.ac.uk, \quad eleanor.k.harris@kcl.ac.uk}}
\end{center}

\vspace{4mm}
 
\vspace*{0.6cm}

%\end{center}
\vspace*{1.5cm}
\begin{abstract}
\noindent We investigate dilaton-gravity models whose solutions  contain a large portion of the static patch of dS$_2$. The thermodynamic properties of these theories are considered both in the presence of a finite Dirichlet wall, as well as for asymptotically near-AdS$_2$ boundaries. We show that under certain circumstances such geometries, including those endowed with an asymptotically near-AdS$_2$ boundary, can be locally and even globally thermodynamically stable within particular temperature regimes. First order phase transitions reminiscent of the Hawking-Page transition are discussed.  For judiciously chosen models, the near-AdS$_2$ boundary can be viewed as a completion of the stretched cosmological dS$_2$ horizon. We speculate on candidate microphysical models.
%can be endowed with an asymptotically near-AdS$_2$ boundary and 
 
\end{abstract}

\newpage
\setcounter{page}{1}
\pagenumbering{arabic}

\tableofcontents

\section{Introduction}%, which approximately describes such an expanding spacetime
Given the measured accelerated expansion of our universe, one  hopes that an underlying microscopic theory exists which describes the quantum properties of de Sitter.  In the spirit of \cite{Farhi:1989yr,Freivogel:2005qh}, a possible avenue to try and define such a theory could be to embed a piece of $(d+1)$-dimensional de Sitter within a $(d+1)$-dimensional  anti-de Sitter geometry, and then use the tools of the AdS/CFT correspondence to capture the physics of this internal, expanding region. In $d\ge 3$, it has been shown that such a set-up either defies the null-energy condition or that this geometry can only be realised if the de Sitter region is surrounded by a black hole horizon \cite{Farhi:1989yr,Freivogel:2005qh}, making it difficult to probe its microscopic structure. However, in two spacetime dimensions geometries embedding dS$_2$ within AdS$_2$ have been constructed which moreover obey the null-energy condition when viewed as pieces of higher-dimensional solutions \cite{Anninos:2020cwo}. We will refer to these as interpolating geometries \cite{Anninos:2017hhn,anninos2019sitter, Chapman:2021eyy}. They offer the possibility of a holographic interpretation for the static patch of de Sitter. The idea of rearranging the microphysical Hilbert space of quantum gravity in AdS to obtain dS microstates also plays an interesting role  in the approach of \cite{Gorbenko:2018oov,Coleman:2021nor,Shyam:2021ciy}, and \cite{Susskind:2021esx,Ecker:2022vkr}.
\\ \\
A particular way interpolating geometries containing a region of dS$_2$ embedded in an AdS$_2$ world can arise is as part of the solution space of dilaton-gravity models for certain choices of the dilaton potential \cite{Anninos:2017hhn,Witten:2020ert}. The solutions permit an asymptotically near-AdS$_2$ boundary of the type studied in \cite{Maldacena:2016upp}. When the asymptotically near-AdS$_2$ interpolating geometry ends at a dS$_2$ `black hole' horizon one must grapple with a horizon of negative specific heat. Alternatively, one can choose a dilaton profile that decreases towards the near-AdS$_2$ boundary \cite{Anninos:2017hhn} leading to a dS$_2$ `cosmological' horizon with positive specific heat, but now at the cost of a non-standard boundary behaviour for the dilaton. Neither of these features is an insurmountable obstacle in the search of a microphysical completion, but each adds an additional layer of difficulty. 
\\ \\
In this paper, we explore mechanisms for stabilising a portion of the dS$_2$ static patch within a near-AdS$_2$ world with ordinary boundary behaviour for the dilaton. We study Dirichlet boundary conditions on the geometry at a  finite proper distance from the horizon, following the method of York \cite{York:1986it,Whiting:1988qr} applied to the de Sitter case \cite{Hayward:1990zm,Wang:2001gt,Anninos:2011zn,Banihashemi:2022jys,Banihashemi:2022htw}.\footnote{Ordinary JT gravity with Dirichlet boundary conditions was considered, for example, in \cite{Gross:2019ach,Gross:2019uxi,Iliesiu:2020zld,stanfordJT, Griguolo:2021wgy}.} The temperature is fixed to be the Tolman temperature \cite{Tolman:1930zza}, which is the proper length of the Euclidean boundary $S^1$. Moreover, we construct interpolating geometries which include an additional piece of the Euclidean AdS$_2$ in the deep interior, leading to asymptotically near-AdS$_2$ solutions with positive specific heat and a large portion of Euclidean dS$_2$ in the interior.%, isolating the static patch of de Sitter inside Anti-de Sitter  \\
\\ \\
As a warm up, we will first consider the thermodynamics of the dimensional reduction of the near-Nariai solution along the lines of \cite{Anninos:2017hhn,anninos2019sitter,Svesko:2022txo}. Throughout, we will be in Euclidean signature, in which the near-Nariai geometry analytically continues to the round $S^2$ with a running dilaton. The poles of the two-sphere give the positions of the black hole horizon, and the cosmological horizon. Because the geometry caps off in this way, it is a natural setting to ask about the effects of a Dirichlet boundary cutting the two-sphere. When we select the solution for which the dilaton grows towards the boundary, and away from the horizon, the piece of the geometry containing the `black hole' dS$_2$ horizon is thermodynamically stable. The piece containing the `cosmological' dS$_2$ horizon, which has a decreasing dilaton profile, is thermodynamically unstable. 
\\ \\
We  then consider the Dirichlet problem for the interpolating geometry, with standard boundary behaviour for the dilaton.  In this case,  thermal stability of the dS$_2$ horizon depends on the location of the Dirichlet boundary and the boundary value of the dilaton field. When the boundary is well approximated by that of an asymptotically near-AdS$_2$ geometry, the specific heat of the dS$_2$ horizon is negative. As the Dirichlet boundary is brought in, along with a tuning of the boundary value of the dilaton, a transition occurs rendering the specific heat positive. To preserve a locally thermally stable interpolating solution that is asymptotically near-AdS$_2$, with a portion of Euclidean de Sitter in the interior, we must add a further deformation of the geometry near the `black hole' dS$_2$ horizon. We provide explicit constructions that accomplish this. Although the solution removes both Euclidean horizons of dS$_2$, we note that we can tune the dilaton potential in such a way that the dS$_2$ portion has a region that is parameterically close to one with a cosmological horizon, while preserving local stability. A similar excision is considered in the stretched horizon picture \cite{Susskind:1993if}, whereby one considers physics on a timelike hypersurface parameterically near the event horizon. In Euclidean signature, this excision corresponds to removing a small disk surrounding the Euclidean horizon. As such, we end up with a geometry that is both thermally stable and encodes a stretched `cosmological' dS$_2$ horizon, while being asymptotically near-AdS$_2$. 
%  {\color{blue}xxx}
\\ \\
In section \ref{dilgravsec}, we introduce a general class of Euclidean dilaton-gravity theories and provide general formulae for their thermodynamic properties in the presence of both a Dirichlet boundary at finite proper distance, as well as for an asymptotically AdS$_2$ boundary.  In section \ref{Near-Nariai geometry}, we utilise these formulae in the near-Nariai limit of the black hole in dS$_2$ which in Euclidean signature is an $S^2$. We show, as in \cite{Svesko:2022txo}, that the black hole is thermodynamically stable, while the cosmological horizon is thermodynamically unstable in this theory. In section \ref{(A)dS$_2$ Interpolating geometries}, we study the Dirichlet problem for the interpolating geometry and show that this saddle may be both stable, and thermodynamically favoured, depending on where we place the Dirichlet wall. We also posit that a region within this geometry may be interpreted as a completion of the stretched dS$_2$ cosmological horizon to a near-AdS$_2$ boundary. In section \ref{disec}, we introduce a dilaton-gravity theory permitting a solution which  interpolates to a dS$_2$ region while being AdS$_2$ in the deep interior as well as near the boundary. We show that this saddle has positive specific heat, while also having the virtue of being asymptotically AdS$_2$. First order phase transitions, reminiscent of the Hawking-Page transition \cite{Hawking:1982dh}, between the interpolating saddle and the AdS$_2$ black hole are discussed. Finally, in section \ref{outlooksec}, we  discuss the possibility of realising a boundary matrix model or SYK-type model that may capture some features of the expanding piece of the geometry.  Appendices \ref{appAdS2} and \ref{Double Interpolating Geometry finite} summarise the thermodynamic properties of some further examples of dilaton potentials with finite boundaries.

\section{Thermodynamics of two-dimensional dilaton-gravity}\label{dilgravsec}

In this section, we introduce the general class of models we will be interested in. These are two-dimensional dilaton-gravity models whose field content is given by a two-dimensional metric, $g_{\mu\nu}$, and the dilaton field, $\phi$. The Euclidean action governing the theory is given by
\begin{equation} \label{dilatongravity}
    S_E = S_0 -\frac{1}{2} \int_{\mathcal{M}} d^2x \sqrt{g} ( \phi R + V(\phi)) - \int_{\partial \mathcal{M}} d \tau \sqrt{h} K \phi_b~,
\end{equation}
where the dilaton potential $V(\phi)$ will be a general function of $\phi$ unless otherwise specified. The theory is considered on a disk topology $\mathcal{M}$ with boundary $\partial \mathcal{M} = S^1$. The trace of the extrinsic curvature normal to $\partial \mathcal{M}$ is given by $K$, and $\phi_b \equiv \phi|_{\partial \mathcal{M}}$ is the boundary value of the dilaton. The first term in (\ref{dilatongravity}) is the contribution from the constant part of the dilaton
\begin{equation}
	S_0 = - \frac{\phi_0}{2} \int_{\mathcal{M}} d^2 x \sqrt{g} R - \phi_0 \int_{\partial \mathcal{M}} d\tau \sqrt{h} K  = -2\pi \phi_0 \chi~.
\end{equation}
Due to the Gauss-Bonnet theorem, this term computes the Euler character $\chi$ of $\mathcal{M}$ and so is topological. For a disk topology one has $\chi = 1$. In what follows, will be interested in the canonical partition function for arbitrary $V(\phi)$ in the presence of a Euclidean boundary at finite proper distance from the Euclidean horizon. 

\subsection{Equations of motion and solutions}
The equations of motion stemming from (\ref{dilatongravity}) are
\begin{align}
    \nabla_\mu \nabla_\nu \phi - g_{\mu \nu} \laplacian \phi + \frac{1}{2} g_{\mu \nu} V(\phi) &= 0~,  \label{eom1}\\
    R + \partial_\phi V(\phi)&= 0~. \label{Riccieom}
\end{align}
These equations can be rearranged, at least within a local neighbourhood of $\mathcal{M}$, as follows
\begin{align}
 - \laplacian \phi + V(\phi) &= 0~, \label{EOM1}\\
    R + \partial_\phi V(\phi)&= 0~, \label{EOM2} \\ 
    \nabla_\mu \xi_\nu + \nabla_\nu \xi_\mu &= 0~, \quad \xi^\mu \equiv \varepsilon^{\mu\nu} \partial_\nu \phi~. \label{EOM3}
\end{align}
Therefore, even though it seems that (\ref{eom1}) and (\ref{Riccieom}) are overconstrained, recasting them in the form (\ref{EOM1}), (\ref{EOM2}), and (\ref{EOM3}) shows that there are only two second order equations acting on the degrees of freedom $\phi$ and $R$. The equation (\ref{EOM3}) moreover indicates that $\xi^\mu$ is a Killing vector field of $g_{\mu\nu}$. Using the existence of $\xi^\mu$ and the two diffeomorphisms we can place a general solution of  (\ref{dilatongravity}) in the form
\begin{equation} \label{generalsolution}
    ds^2 = N(r) d\tau^2 + \frac{dr^2}{N(r)}~, \qquad \phi(r) =  r~, \qquad \tau \sim \tau+\beta~.
\end{equation}
%In deriving a dilaton-gravity theory by dimensional reduction from a four dimensional space such as the Reissner-Nordstr\"{o}m black hole, the dilaton appears as the area of the two-sphere. Therefore, it is sensible to choose the dilaton to grow in the direction of the radius.
The Ricci scalar of this metric is $R = - N''(r)$. This combined with the equation of motion (\ref{Riccieom}) gives 
\begin{equation}
    \partial_\phi V(\phi) = N''(r) ~,
\end{equation}
which gives the following relation between the dilaton potential and the coefficient of the metric
\begin{equation} \label{V-Areleation}
    N(r, r_h) = \int_{r_h}^r dr' \, V(r') >  0  ~.
\end{equation}
Since we are considering Euclidean solutions, this must be positive for all $r$. Noting that $N(r_h,r_h) = 0$, the metric (\ref{generalsolution}) describes a Euclidean black hole solution with the horizon at $r_h$, and where $\phi(r_h)  \equiv \phi_h$ is the value of the dilaton at the horizon. We emphasise that $r_h$ is not an independent variable as it is fixed by the form of $N(r,r_h)$. Expanding (\ref{generalsolution}) near to the horizon $r = r_h$ and imposing periodicity of $\tau \sim \tau +\beta$ to ensure smoothness of the Euclidean solution leads to the relation
\begin{equation} \label{beta}
    \beta = \frac{4\pi}{|V(r_h)|}~.
\end{equation}
This formula holds regardless of the asymptotic form of the potential. The usual JT gravity \cite{Jackiw:1984je,JT} corresponds to the case where $V(\phi) = 2 \phi$. For this potential (\ref{Riccieom}) ensures that the curvature is constant and negative, giving an AdS$_2$ solution for the metric. 

\subsection{Boundary conditions} \label{bcsec}

As boundary conditions, we consider the Euclidean Dirichlet problem for which the  induced metric $h$ at $\partial\mathcal{M}$ and the corresponding proper length $\beta_T$, as well as the boundary value of the dilaton $\phi_b$ on $\partial \mathcal{M}$ are fixed. The gauge parameter $\xi_\mu$ is required to vanish at $\partial\mathcal{M}$, to ensure the boundary condition on $h$ is preserved and the location of the boundary is not disrupted (see for instance \cite{Witten:2018lgb}).
\newline\newline
The induced metric $h$ can be arranged to be constant along $\partial\mathcal{M}$ by a judicious reparameterisation of the boundary time $\tau$. In what follows, we moreover take $\phi_b$ to be constant along $\partial \mathcal{M}$. The physical motivation behind this choice of boundary condition stems from the Gibbons-Hawking prescription for Euclidean black hole thermodynamics \cite{Gibbons:1976ue,Gibbons:1977mu}, whereby one views the Euclidean solutions on the disk as contributing to the thermal partition function of the underlying theory.\footnote{Alternative boundary conditions could fix the trace of the extrinsic curvature $K$ at $\partial\mathcal{M}$ or the normal derivative of $\phi$ at $\partial\mathcal{M}$, or more general combinations thereof.} We will not necessitate that the boundary is asymptotically (near) AdS$_2$  for large part of our discussion, in line with the considerations of \cite{Svesko:2022txo}. \\
\\
Given the solution (\ref{generalsolution}) and (\ref{V-Areleation}), it is clear that a general choice of $\phi_b$ and $h$ will not permit a  real solution to our boundary value problem. For instance, let us consider the dilaton potential with $V(\phi) = -2\phi$. This yields   $N(r,r_h) = r_h^2- r^2$ and $\phi(r) = r$. Smoothness of the solution at the origin of the disk moreover fixes $r_h = 2\pi/\beta$ such that the proper size of the boundary circle is $\beta_T = \sqrt{4\pi^2-\beta^2 r_b^2}$. Regardless of our choice for $\phi_b$ and assuming our solution is real, we have that $\beta_T \in (0,2\pi)$ which is only a subset of the positive half-line. More markedly, for $\phi_b=0$, only $\beta_T = 2\pi$ is permitted. In particular, going from a non-vanishing to a vanishing value of $\phi_b$ leads to a discontinuous jump in the allowed set of $\beta_T$. These restrictions on independent boundary data are a two-dimensional analogue of the obstructions arising when setting up the Dirichlet problem in four-dimensional general relativity \cite{Andrade:2015gja,An:2021fcq,Anderson:2007jpe,Witten:2018lgb} on manifolds with a boundary. In the case at hand, aside from such restrictions, the Dirichlet problem is well-posed due to the absence of locally propagating gravitational degrees of freedom. Nonetheless, we must ensure the Dirichlet data $(\beta_T,\phi_b)$ we consider are indeed sensible.

\subsection{Thermodynamics} \label{Thermodynamics}

%footnote{The thermodynamic properties of the four-dimensional Schwarzschild solution placed in a finite sized box were originally studied in \cite{York:1986it}.} 
We would like to study the thermodynamics of the class of theories (\ref{dilatongravity}) in the presence of  a finite boundary. One motivation for studying the finite boundary case is that certain geometries, such as the static patch of de Sitter space, do not afford an asymptotic boundary. In such circumstances, it is natural to consider a Dirichlet problem with a finite boundary. Our goal here is to present formulae for thermodynamic quantities at a finite boundary, generalising the results of \cite{Anninos:2017hhn,Grumiller:2007ju,Witten:2020ert}. \\
%the Nariai solution, which describes the near horizon region of the Schwarzschild-de Sitter black hole and was studied in \cite{}. 
%In the Euclidean continuation of the Nariai solution, the geometry caps off at both the black hole and cosmological horizon, describing a two-sphere. 
\\
We now evaluate the on-shell Euclidean action, which is related to the partition function by $-S_E = \log Z$. Given the Dirichlet problem under consideration, we have that $Z$ is a function of $\phi_b$ and $\beta_T$. The induced metric $h$ and $\phi_b$ at $\partial\mathcal{M}$ are given by
\begin{equation} \label{boundarymetric}
    ds^2_b = N(r_b ,r_h) d\tau^2 \equiv h \, d\tau^2~, \quad\quad \beta_T = \beta \sqrt{N(r_b,r_h)}~, \quad\quad \phi_b = r_b~.
\end{equation}
The trace of the extrinsic curvature is  
\begin{equation}
    K = \frac{1}{h} K_{\tau \tau} = \frac{1}{2\sqrt{h}} \, \partial_{r_b} N(r_b,r_h)~.
\end{equation}
We will further assume, for now, that the dilaton takes its minimal value at the Euclidean black hole horizon $r_h$. This is not necessarily the case, for example, when considering cosmological horizons the horizon will be located at the largest possible value of the dilaton, as will discussed in the next section. The on-shell action is  
\begin{equation}
    \begin{split}
        S_E &=  \frac{\phi_0 \beta}{2} \int_{r_h}^{r_b} dr  N''(r) - \frac{\phi_0 \beta}{2} N'(r_b) - \frac{\beta}{2} \int_{r_h}^{r_b} dr \left( -  r N''(r) +   N'(r) \right) - \frac{\beta}{2}N'(r_b) \phi_b~.
    \end{split}
\end{equation}
Integrating the first term in the bulk action by parts and using (\ref{beta}), along with the fact that for a black hole horizon $V(r_h)>0$ (otherwise (\ref{V-Areleation}) would not hold for $r$ slightly greater than $r_h$), this evaluates to 
%\begin{equation} \label{Onshellresult}
%     S_E = - 2 \pi \text{Sgn}[A'(r_h)] ( \phi_0 + \phi_h ) - \beta  A(r_b) + \frac{\beta}{2} r_b V(r_b) - \frac{\beta}{2 } V(r_b) \phi_b  .
%\end{equation} 
\begin{equation} \label{Onshellresult}
  \log Z =  2 \pi\, ( \phi_0 + \phi_h ) + \beta_T  \sqrt{N(r_b,r_h)}~.% + \frac{\beta}{2} r_b V(r_b) - \frac{\beta}{2 } V(r_b) \phi_b  .
\end{equation}
We see that the  expression (\ref{Onshellresult}) takes the suggestive form $\log Z = S- E/T$ for a canonical thermal partition function.  Indeed,   we are interested in the thermodynamics experienced by an observer at a finite cutoff $r_b$. As such, the relevant temperature is the Tolman temperature \cite{York:1986it} which is precisely $\beta_T$ in (\ref{boundarymetric}).
%. The inverse Tolman temperature is defined to be 
%\begin{equation}
%    \beta_T(r_b) = \beta \sqrt{A(r_b)},
%\end{equation}
%where $\beta$ is the inverse Gibbons-Hawking temperature (\ref{beta}). 
To evaluate thermodynamic quantities, we must vary $\log Z$ with respect  to $\beta_T$ while keeping $\phi_b$ (and hence $r_b$) fixed.
%, we fix $r_h$ with respect to $\beta_T$ but allow $A(r_b)$ to vary where previously it was fixed. 
The following expressions prove to be useful
\begin{equation}
    %\begin{split}
        {\partial_{\beta_T} N(r_b, r_h)} = \frac{2 V(r_h)^2 N(r_b, r_h)}{\beta_T( V(r_h)^2 + 2 V'(r_h) N(r_b , r_h) )}~,  \quad\quad  {\partial_{r_h} N(r_b, r_h)}  =  - V(r_h) ~.
    %\end{split}
\end{equation}
%which allow us to define the following thermodynamic quantities: 
%\paragraph{Energy}
%\begin{equation}
%    \begin{split}
%        E = - a \sqrt{A(r_b)}  + \frac{\sqrt{A(r_b)} A''(r_b)}{ A'(r_b)} (a r_b - \phi_b ) . 
%    \end{split}
%\end{equation}
%\paragraph{Entropy}
%\begin{equation}
%    S = 2 \pi \text{Sgn}[A'(r_h)] ( \phi_0 + a r_h )  + \left( \frac{\beta_T \sqrt{A(r_b)} A''(r_b)}{A'(r_b)} - \frac{\beta_T A'(r_b)}{2 \sqrt{A(r_b)}}\right)(a r_b - \phi_b). 
%\end{equation}
%\paragraph{Heat capacity}
%\begin{multline}
%    C = - \beta_T \sqrt{A(r_b)} \bigg[ - a + \big( A''(r_b) A'(r_b)^2 + 2 A(r_b) A'(r_b) A'''(r_b)
%    - 2 A(r_b) A''(r_b)^2 \big)  \frac{(ar_b - \phi_b)}{A'(r_b)^3} \\+ \frac{2a A(r_b) A''(r_b)}{A'(r_b)^2} \bigg]. 
%\end{multline}
The thermal energy $E$, and entropy $S$ are found to be
%and specific heat $C$ are found to be
%If we set $\phi_b = a r_b$ and $\phi_h = a r_h$, these are
%\paragraph{Energy}
\begin{eqnarray}
        E &=& -  \sqrt{N(r_b,r_h)}~, \label{energy}\\ 
         S &=& 2 \pi  ( \phi_0 + \phi_h)~, \label{entropy}
\end{eqnarray}
while the specific heat is given by\footnote{In the case where the dilaton decreases towards the AdS$_2$ boundary, rather than growing, these formulae will be modified to have an overall minus sign.}
\begin{equation}
  C =  \frac{ \beta_T \sqrt{N(r_b,r_h)} \, V(r_h)^2}{V(r_h)^2 + 2 V'(r_h) N(r_b , r_h)}~, \label{heatcapacity}
\end{equation}
in agreement with \cite{Grumiller:2007ju}. In the case where there are multiple possible saddles for a given $\beta_T$, the above formulae will hold at each saddle and the free energy must be computed to determine which solution is dominant. We observe that the entropy (\ref{entropy}) is a function of quantities located at the Euclidean horizon, reflecting its more universal nature. This property was already observed in the early work of York \cite{York:1986it} when placing a four-dimensional black hole in a box with Dirichlet conditions. The energy and specific heat depend more sensitively on the choice of boundary conditions.  Consequently, though the entropy is insensitive to the location of the boundary, the thermal stability of a horizon may depend on this. \\
\\
A special case of the above is when the dilaton potential at large values of $\phi$ takes the form $V(\phi) = 2 \phi +\mathcal{O}(\phi^{-\epsilon})$. Here, our geometries acquire an asymptotic near-AdS$_2$ boundary such that the metric takes the asymptotic form $N(r_b,r_h) \approx r_b^2 - b$, and it follows from (\ref{boundarymetric}) that our thermodynamic quantities are computed (after a rescaling of $r_b$) with respect to the inverse temperature $\beta = 4\pi/V(\phi_h)$. Hence, for asymptotically near-AdS$_2$ configurations, our thermodynamic quantities are more closely associated to the behaviour of the spacetime at the horizon. Specifically, taking $\phi_b \gg 1$ while keeping $\beta_T$ fixed, the quantities (\ref{energy}), (\ref{entropy}), and (\ref{heatcapacity}) become
\begin{equation} \label{asympESC}
 E_\infty =-\phi_b + \frac{b}{2 \phi_b}~,  \quad S_\infty = 2 \pi \left(  \phi_0 +   \phi_h \right)~, \quad C_\infty =-  \frac{2 \pi \, V(\phi_h)}{R(\phi_h)}~,
\end{equation}
in agreement with the results of \cite{Witten:2020ert,Anninos:2017hhn,Grumiller:2007ju}. Stated otherwise, the more rigid nature of the near-AdS$_2$ boundary has allowed for the thermodynamic derivatives to occur with respect to the ordinary Gibbons-Hawking temperature $\beta$ in (\ref{beta}), which is defined as the surface gravity at the horizon with a suitably normalised timelike Killing vector $\partial_\tau$. 
\newline\newline
We now explore the thermodynamic properties of the dimensional reduction of the near-Nariai black hole in the presence of a finite boundary. 

\section{Near-Nariai thermodynamics} \label{Near-Nariai geometry}

In this section, we compute thermodynamic properties of the dimensionally reduced near-Nariai geometry which describes the near horizon limit of the Schwarzschild-de Sitter spacetime with near coincident horizons. The dilaton potential for this model is given by $V(\phi) = -2\phi$. The thermodynamics of this dilaton-gravity model were also discussed in \cite{Anninos:2017hhn,Svesko:2022txo}.

\subsection{Near-Nariai geometry} 

\begin{figure}[H]
	\centering
	\includegraphics[width=10cm]{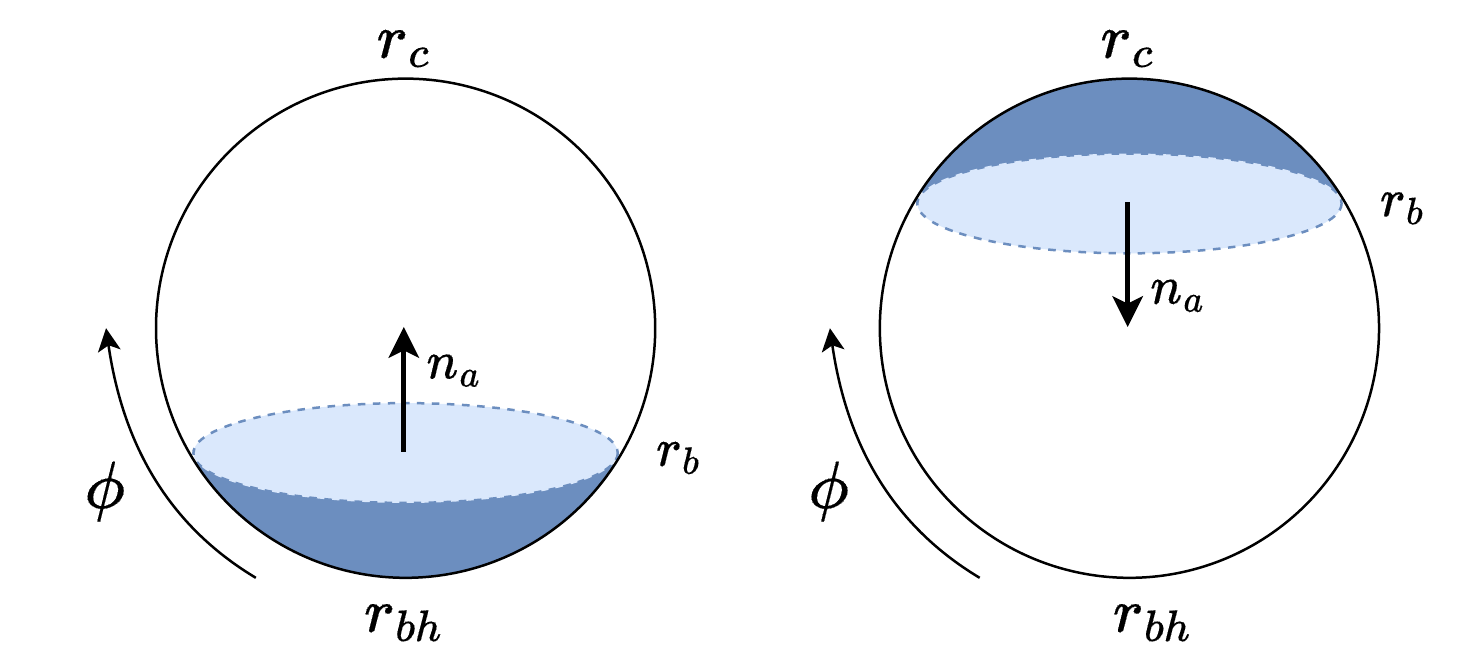}
	\caption{The Euclidean Nariai geometry is an $S^2$. The choice of sign of the outward pointing normal vector is dependent on which horizon we choose to integrate from. The direction of increasing $\phi$ is indicated which we will always take to increase from the back hole horizon towards the cosmological horizon.}
	\label{fig:UnitNormal2}
\end{figure}

\begin{figure}[H]
	\centering
	\includegraphics[width=10cm]{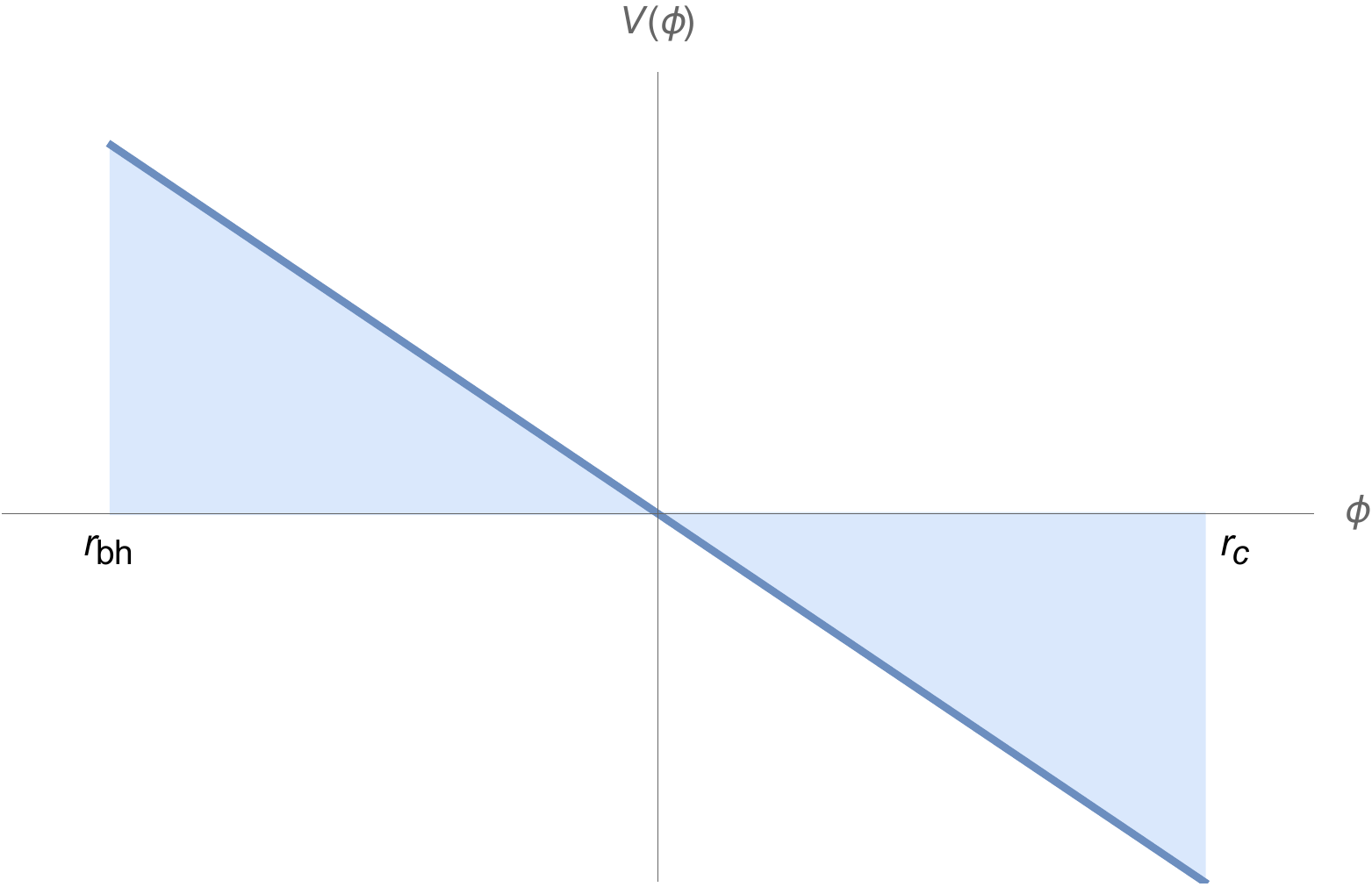}
	\caption{The dilaton potential $V(\phi) = - 2 \phi$ for the black hole in de Sitter. The solution caps off at the two horizons which have been indictated here as $r_{bh}$ and $r_c$. }
	\label{fig:dSBH-V}
\end{figure}

As in the previous section, we will consider the metric in Euclidean signature. The solution reads
\begin{equation} \label{dS2metric}
    ds^2 = (r_h^2 - r^2)d\tau^2 + \frac{dr^2}{(r_h^2 - r^2)}~, \quad \tau \sim \tau + \beta~, \quad\quad  \phi(r) =r~,
    \end{equation}
where $r_h \leq r \leq -r_h$, with $r_h<0$ and $\tau \sim \tau + \beta$. There are two Euclidean horizons at $r = \pm r_h$ -- one is the black hole horizon which we will refer to as $r_{bh} = r_{h} < 0$, and the other is the cosmological horizon which we call $r_c = - r_{h} > 0$. In Euclidean signature the space caps off at the two horizons and so this metric coincides with the round metric on the two-sphere, as illustrated in Figure \ref{fig:UnitNormal2}. Taking $\tau \to i t$ in (\ref{dS2metric}) we retrieve the Lorentzian metric for the static patch of dS$_2$. Since the dilaton is also running, it is a near-dS$_2$ static patch.
As in the previous section, $\beta$ is chosen to ensure smoothness of the Euclidean saddle. Now that there are two such potential singularities, we must have 
\begin{equation} \label{Nariaitemp}
    \beta = \frac{4\pi}{|V(r_{bh})|} = \frac{4\pi}{|V(r_{c})|} =  \frac{2\pi}{|r_h|}~.
\end{equation}
Employing (\ref{V-Areleation}), we can confirm that the dilaton potential that produces this geometry is $V(\phi) = - 2 \phi$, and this potential is plotted in Figure \ref{fig:dSBH-V}. From the plot we see that the condition (\ref{Nariaitemp}) is satisfied since $r_{bh} = -r_{c} = r_h$. We therefore have a $\mathbb{Z}_2$ symmetry $r \rightarrow -r$ in this metric which exchanges the two horizons. It is the sign of the dilaton that decides for us which is the cosmological horizon and which is the black hole, and here we choose the dilaton to always grow away from the black hole horizon. 

\subsection{Near-Nariai with a boundary} 

In order to compute the thermal partition function, we place a boundary at some value $r=r_b \in (r_h,- r_h)$, where $\phi_b = r_b$ and $\beta_T$ are fixed. The boundary can either surround the black hole or cosmological horizon. In fact, there is no choice of saddle here. Rather, all saddles must be included provided they satisfy the same Dirichlet boundary conditions. However, of all the saddles, one will generally dominate. 
%It is also convenient to introduce $\phi_h \equiv r_h$. 
\newline\newline
Thus, we must evaluate the thermodynamic quantities for both horizons in the presence of a boundary at $r_b$. For the black hole horizon at $r=r_h$, the analysis of the previous section can be directly applied, resulting in
\begin{equation} \label{bhonshell}
    \log Z_{\text{bh}} = 2 \pi ( {\phi}_0 + r_h ) + \beta_T \sqrt{r_h^2-\phi_b^2}~, \quad\quad r_h  = -   \frac{2\pi |\phi_b|}{\sqrt{4\pi^2-  \beta_T^2}}~. % {\color{blue}  \text{Sgn}[A'(r_h)] } 
\end{equation}
\newline
 An important difference for the cosmological horizon is that the outward pointing unit normal vector has the opposite sign to that of the black hole case, as shown in Figure \ref{fig:UnitNormal2}. For this reason, the sign of the boundary term must be flipped in order to have a well-posed variational problem. Furthermore, for a cosmological horizon, (\ref{V-Areleation}) becomes 
\begin{eqnarray}
	\int_{r}^{r_c} dr' \, V(r') < 0 ~. 
\end{eqnarray}
So, $V(r_c) < 0$ in order for this to hold for $r$ slightly smaller than $r_c$. %and therefore (\ref{beta}) gives $\beta = 2 \pi / r_c$.  
%Therefore, the Euclidean action for the cosmological horizon is 
%\begin{multline}
%    S_E =  \frac{{\phi}_0}{2} \int_{r_b}^{r_c} dr  \int_{0}^{\beta} d\tau \sqrt{g} R - {\phi}_0 \int_{0}^{\beta} d\tau \sqrt{h} K \\
%    -\frac{1}{2} \int_{r_b}^{r_c} dr \int_{0}^{\beta} d\tau \sqrt{g} ( \phi R + V(\phi)) + \int_{0}^{\beta} d \tau \sqrt{h} K \phi_b~.
%\end{multline}
%This change in the sign of the outward pointing normal vector is counteracted by the fact that we have to flip the limits of integration as now $r_b \leq r_c$. 
The end result is a change in the sign of the second term in equation (\ref{Onshellresult}), such that
\begin{equation} \label{cosmoonshell}
    \log Z_{\text{cos}} = 2 \pi ( {\phi}_0 - r_h ) - \beta_T \sqrt{r_h^2-\phi_b^2}~.%, \quad\quad r_h  = - \frac{2\pi |\phi_b|}{\sqrt{4\pi^2-  \beta_T^2}}~. % {\color{blue}  \text{Sgn}[A'(r_h)] } 
\end{equation}
%For the theory to have well defined boundedness properties one must ensure that ${\phi}_0 - \phi \gg 1$. 
We can now compute and compare various thermodynamic properties. The canonical free energies of the two saddles are
\begin{eqnarray}
F_{\text{bh}} &=&   - \frac{2 \pi  {\phi_0}}{\beta_T} +  \frac{|\phi_b|\sqrt{4 \pi ^2- \beta_T^2}}{\beta_T} ~, \\%-\beta_T^{-1} \left( 2 \pi {\color{blue}   } ( {\phi}_0 - r_h ) + \beta_T \sqrt{r_h^2-r_b^2} \right)~, \\ 
F_{\text{cos}} &=& - \frac{2 \pi  {\phi_0}}{\beta_T} -\frac{|\phi_b|\sqrt{4 \pi ^2- \beta_T^2}}{\beta_T} ~.
\end{eqnarray} 
It follows that
\begin{equation}
\Delta F \equiv F_{\text{bh}} - F_{\text{cos}} = \frac{2  |\phi_b|}{\beta_T} \sqrt{4\pi^2 - \beta_T^2} \ge 0~,
\end{equation}
where the inequality is saturated for $\phi_b = 0$.\\
\\
The saddle with lowest free energy is the one with the cosmological horizon. However, we must also assess whether the saddle is stable under small thermal fluctuations. We thus compute the specific heat
\begin{equation}\label{Cnariai}
C_{\text{bh}} = - C_{\text{cos}} = %\frac{\phi_h^2 \beta_T}{ \phi_b^2} \sqrt{\phi_h^2 - \phi_b^2} = 
\frac{4 \pi^2 |\phi_b| \beta_T ^2}{\left(4 \pi^2-\beta_T^2\right)^{3/2}}> 0~,
%C_{\text{cos}} &=& -\frac{\phi_h^2 \beta_T}{ \phi_b^2} \sqrt{\phi_h^2 - \phi_b^2} < 0~.
\end{equation}
finding that with Dirichlet boundary conditions, the black hole horizon of the near-Nariai geometry has positive specific heat while the cosmological horizon has negative specific heat \cite{Svesko:2022txo} .
%\footnote{Near $\phi_b = 0$, we can define as scaling limit $\phi_b = \varepsilon \tilde{\phi}_b$ and $\beta_T = 2\pi+ \varepsilon^{2/3} \tilde{\beta}_T$, and take the limit $\varepsilon \to 0^-$. } 
This also resonates with the results of \cite{Draper:2022ofa,Banihashemi:2022htw}. In consequence, even though the cosmological saddle has lower free energy, it is thermally unstable in the setup under consideration. At least locally, the black hole saddle is the stable one, although it may be metastable at the non-perturbative level. \\
\\
The energy can be computed in a similar fashion via  (\ref{energy}). 
We find
%paragraph{Black hole horizon:} 
%Using equation, the average energy for the piece of the geometry containing the black hole is 
\begin{equation}
    E_{\text{bh}}  = -E_{\text{cos}} = -  \frac{|\phi_b| \beta_T}{\sqrt{4\pi^2 - \beta_T^2}}~.
    %, \\ E_{\text{cos}} &=& \sqrt{\phi_h^2 - \phi_b^2}~. 
\end{equation}
The energies of the black hole and cosmological horizons are equal and opposite. %We note that the energy of the black hole is negative but bounded below with ground state energy $-A(r_b, r_h)$. 
%\paragraph{Cosmological horizon:} For the cosmological horizon the difference in the sign of the on-shell action compared to that of the black hole again results in an overall difference of sign in the energy: 
%\subsubsection*{Entropy}
We can similarly calculate the entropy by either integrating from the black hole horizon, as in the left hand side of Figure \ref{fig:UnitNormal2} or the cosmological one as in the right hand side of this figure. Using equation (\ref{entropy}), we have
\begin{eqnarray}
    S_{\text{bh}} &=& 2 \pi \left( \phi_0 +  \phi_{h} \right)~, \\
      S_{\text{cos}} &=& 2 \pi \left( \phi_0 -  \phi_{h} \right)~,
\end{eqnarray}
%\paragraph{Cosmological horizon:} Again, the overall sign difference of the on-shell action requires us to take the negative of  equation (\ref{entropy}) to be the entropy of the cosmological horizon, but this is counteracted by the fact that $V(\phi_c) <0$. Therefore, the entropy in this case is  
where $\phi_h  = r_h <0$ is the value of the dilaton at the horizon. \\
\\
It is interesting to note that the total energy in the near-dS$_2$ static patch, which is given by `gluing' the black hole saddle to the cosmological saddle along their common boundary with $\phi_b =0$, yields a vanishing result
\begin{equation} \label{totalnariaienergy}
    E_{\text{tot}} \equiv E_{\text{bh}}  + E_{\text{cos}}  = 0~. 
\end{equation}
Adding the entropies together yields the total entropy of the near-dS$_2$ static patch to be
\begin{equation} \label{totalnariaientropy}
    S_{\text{tot}} \equiv S_{\text{bh}}  + S_{\text{cos}}  = 4 \pi \phi_0~. 
\end{equation}
We can compare (\ref{totalnariaienergy}) and (\ref{totalnariaientropy}) to the result from calculating the classical contribution to the partition function over the whole two-sphere, where now there is no longer a boundary term for the action. One finds
\begin{equation}
\begin{split}
    S_E %&=  \frac{\phi_0 \beta}{2} \int_{-r_h}^{r_h} dr \, N''(r) - \frac{\beta}{2} \int_{-r_h}^{r_h} dr \, \left[ - r N''(r) + N'(r) \right]   \\%
    &= \frac{\phi_0 \beta}{2} \left( V(-r_h) - V(r_h) \right) - \frac{\beta \, r_h}{2} \left(V(-r_h) + V(r_h)  \right).
\end{split}
\end{equation}
Since for the Nariai geometry $V(r_h) = -2r_h$ the second term vanishes. The free energy $F$ is related to the partition function by $\ln Z = - \beta F = S-\beta E$. Since there is no spatial boundary anymore, the total energy of the full sphere vanishes. We therefore have that the total entropy is
\begin{equation}
    S_{\text{full}} = - S_E = 4 \pi \phi_0~,
\end{equation}
which exactly matches equation (\ref{totalnariaientropy}). It would be good to understand whether this agreement between the full two-sphere and the `glued' saddles persists beyond the classical level by computing quantum corrections as in \cite{Anninos:2021ene,Muhlmann:2022duj, Muhlmann:2021clm}. 

\begin{center} *** \end{center}
%\newline\newline
It is interesting to contrast (\ref{Cnariai}) with  the specific heat of the four-dimensional Nariai black hole, which has been reported in the literature \cite{Niemeyer:2000nq,Anninos:2010gh} to be negative leading to an evaporating black hole (like that of the Schwarzschild black hole in flat space). The four-dimensional thermodynamic quantities rely on a more local treatment of the horizon thermodynamics. The temperature $T$, for instance, is given by the surface gravity at the Killing horizon with a suitably normalised Killing vector. If one computes the area, and hence the Bekenstein-Hawking entropy, of the black hole horizon as a function of $T$, one notes that it decreases with increasing $T$ indicating a negative specific heat. For similar reasons the de Sitter horizon has been reported to have positive specific heat \cite{Anninos:2012qw}. The difference in sign with the computations above is accounted for by the different definition of temperature, which is now given by  the Tolman temperature $\beta_T$. This the near-Nariai black hole version \cite{Svesko:2022txo} of York's observation \cite{York:1986it} that when placed in a box with Dirichlet conditions, the flat space Schwarzschild black hole can have positive specific heat. 
\newline\newline
In the following section we proceed to consider a different treatment of the boundary of the near-Nariai black hole geometry and show that one can recover a negative specific heat.

\section{(A)dS$_2$ Interpolating geometries} \label{(A)dS$_2$ Interpolating geometries}

In this section we explore an extension  \cite{Anninos:2017hhn,anninos2019sitter} of the dS$_2$ dilaton potential, $V(\phi)=-2\phi$, that results in an asymptotically near-AdS$_2$ geometry ending at the near-Nariai black hole horizon. Our goal is to study the thermodynamic properties of such models in the presence of a Dirichlet boundary (a related discussion can be found in \cite{Gross:2019ach}). We end with a comment about the stretched dS$_2$ horizon in this framework. 
%It is possible to choose the dilaton potential $V(\phi)$ in such a way that the geometry interpolates smoothly between a portion of the static patch of dS$_2$ and an asymptotically AdS$_2$ region. These class of theories are known as `centaur' geometries \cite{anninos2019sitter,anninos2018infrared}. 

\subsection{Geometry}

\begin{figure}[H]
	\centering
	\includegraphics[width=10cm]{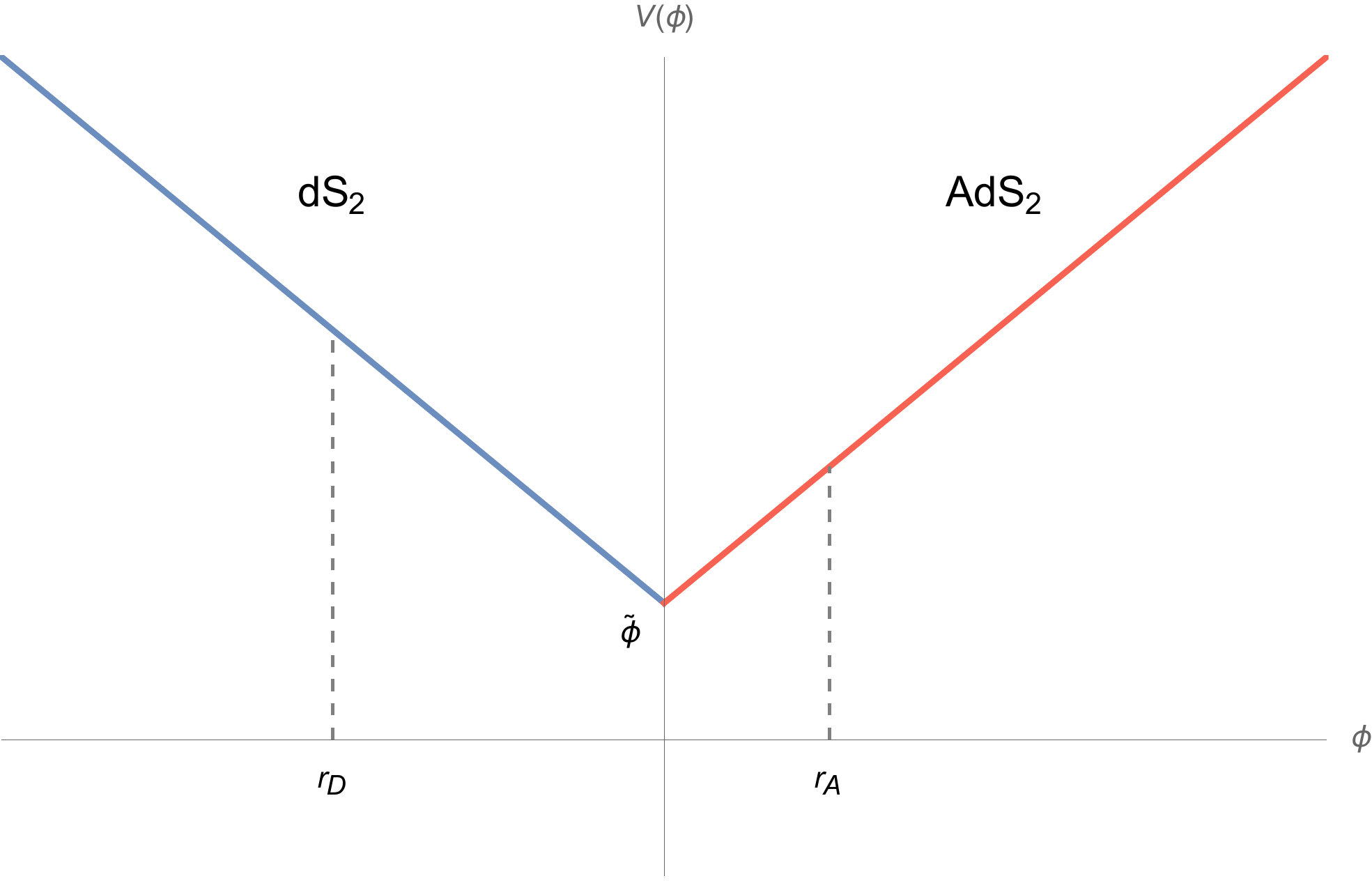}
	\caption{The dilaton potential for the interpolating geometry (\ref{centaurV}) with $\tilde{\phi} >0$. The blue slope indicates where the potential describes a dS$_2$ geometry and the red slope shows where the geometry is Euclidean AdS$_2$. At a given temperature, there can be two saddles, indicated here at $r_D$ and $r_A$.}
	\label{fig:CentaurPotential}
\end{figure}

For the sake of concreteness we focus on the dilaton potential
\begin{equation}\label{centaurV}
    V(\phi) = 2  |\phi  | +  \tilde{\phi}~,
\end{equation}
where $\tilde{\phi}$ is a real-valued parameter.\footnote{It is worth emphasising that any $V(\phi)$, including smooth ones, having regions linear in $\phi$ with oppositely signed slopes and an asymptotic growth of the  near-AdS$_2$ type $V(\phi) = 2\phi + \mathcal{O}(\phi^{-\epsilon})$ suffice for our purposes. A simple example is $V_\varepsilon(\phi) = 2 \phi \tanh \frac{\phi}{\varepsilon}$ with $\varepsilon$ small and positive. Moreover, although we have chosen the slope in  (\ref{centaurV}) to have the same absolute value for $\phi>0$ and $\phi < 0$ one can also consider cases where the slopes differ.} The potential is shown in Figure \ref{fig:CentaurPotential}.  For a given temperature, this geometry will have up to two saddles. We shall refer to these as the interpolating geometry and the AdS$_2$ geometry. Although the potential $V(\phi)$ we study has a jump in its first derivative, the geometries it produces have continuous and differentiable metrics. We now provide the form of the metric for both $\tilde{\phi}$ positive as well as negative.%Moreover, for a potential of the type $V(\phi)$ the on-shell action comes entirely from the boundary term. 

\subsubsection*{Case 1: $\tilde{\phi} \ge 0$}

From equation (\ref{V-Areleation}) we can see that the metric for the interpolating geometry will be of the form (\ref{generalsolution}) with 
\begin{equation} \label{InterpCentaursaddle}
    N^{D_+}(r, r_D) = 
    \begin{cases} 
    (r - r_D)( \tilde{\phi} - r - r_D)~, \qquad & r_D \leq r \leq 0 ~,\\
    r^2 + r_D^2 + \tilde{\phi}(r - r_D)~, \qquad & 0 < r~,
    \end{cases} 
\end{equation}
where $r_D$ defines the Euclidean dS$_2$ black hole horizon. The Euclidean time periodicity is given by
\begin{equation}\label{betaD}
\beta_{D} = \frac{4\pi}{|  \tilde{\phi} -2  r_D |}~.
\end{equation}
We therefore have a portion of the two-sphere (Euclidean dS$_2$) for $r \leq 0$ (in particular, when $\tilde{\phi} = 0$ this is exactly the metric in (\ref{dS2metric})), and when $r > 0$ we have a Euclidean AdS$_2$ metric. Depending on the sign of $\tilde{\phi}$ the two-sphere is glued to a portion of a quotient of the hyperbolic disk ($\tilde{\phi}>0$) or a portion of the hyperbolic strip ($\tilde{\phi} \le 0$). \\
\\
The metric of the AdS$_2$ saddle is 
\begin{equation} \label{AdSCentaursaddle}
	N^{A_+}(r,r_A) = r^2 - r_A^2 + \tilde{\phi}(r - r_A)~, \qquad 0 \leq r_A \leq r  ~,
\end{equation}
where $r_A$ defines the location of the Euclidean AdS$_2$ black hole horizon. The Euclidean time periodicity is given by
\begin{equation}\label{betaA}
\beta_{A} = \frac{4\pi}{|\tilde{\phi} + 2  r_A  |}~.
\end{equation}

\subsubsection*{Case 2: $\tilde{\phi} <0$}

When $\tilde{\phi} < 0$, we must further ensure that the metric is everywhere positive. The AdS$_2$ saddle will remain of the same form, but the range of $r_A$ is modified as
\begin{equation} \label{AdSCentaursaddleNeg}
	N^{A_-}(r, r_A) = r^2 - r_A^2 + \tilde{\phi}(r - r_A)~, \qquad |\tilde{\phi}| /2 \leq r_A \leq r  ~.
\end{equation}
The interpolating solution similarly restricts the range of $r_D$, such that
\begin{equation} \label{InterpCentaursaddleNeg}
    N^{D_-}(r, r_D) = 
    \begin{cases} 
    (r - r_D)( \tilde{\phi} - r - r_D)~, \qquad &  r_D \leq r \leq 0 ~, \qquad r_D < \left(\frac{1+\sqrt{2}}{2} \right) \tilde{\phi}~,\\
    r^2 + r_D^2 + \tilde{\phi}(r - r_D)~, \qquad & 0 < r~.
    \end{cases} 
\end{equation}
In the above we are assuming that the coordinate $r$ can become large compared to $\tilde{\phi}$ and $r_D$. The periodicities $\beta_D$ and $\beta_A$ will be as for $\tilde{\phi}>0$, namely (\ref{betaD}) and (\ref{betaA}) respectively. %{\color{blue}xxx}

\subsection{Thermodynamic properties} \label{Interpolating Thermodynamic properties}

As in the previous sections, we consider Dirichlet boundary conditions whereby the proper size of the boundary circle $\beta_T$ and the boundary value of the dilaton $\phi_b$ are fixed. We take $\phi_b >0$ here since non-positive $\phi_b$ reproduces the near-Nariai black hole setup described in Section \ref{Near-Nariai geometry}. In general, there are two solutions obeying the boundary conditions which we call $r_D$ and $r_A$, as shown in Figure \ref{fig:CentaurPotential}. 
%\subsubsection{Case 1: $\tilde{\phi} >0$}
\newline\newline
The temperatures of the interpolating geometry and the AdS$_2$ geometry are 
\begin{eqnarray}
	\beta_T^D &=& \frac{4\pi}{|\tilde{\phi} - 2 \phi_D|} \sqrt{ r_b^2 + \phi_D^2 + \tilde{\phi}(r_b - \phi_D)} ~, \\
	\beta_T^A &=& \frac{4\pi}{|\tilde{\phi} + 2 \phi_A|} \sqrt{r_b^2 - \phi_A^2 + \tilde{\phi}(r_b - \phi_A)} ~,
\end{eqnarray}
where we have introduced the notation $ \phi_D = r_D$ to be the value of the dilaton at the dS$_2$ black hole horizon and $ \phi_A = r_A$ to be the value of the dilaton at the AdS$_2$ black hole horizon. We must set the temperatures equal to each other $\beta_T^D = \beta_T^A = \beta_T$ to compare their thermodynamic properties at a given temperature. %We further remind the reader that for $\tilde{\phi}<0$ the ranges of $\phi_D$ and $\phi_A$ are restricted.    \\

\subsubsection*{Case 1: $\tilde{\phi} \ge 0$}
Let us begin by taking $\phi_b$ to be large compared to $\tilde{\phi}$ and $\phi_{A,D}$. %r than $\tilde{\phi}$%
The thermodynamics for the Euclidean AdS$_2$ solution (\ref{AdSCentaursaddle}) with $\tilde{\phi}=0$ are reviewed in appendix \ref{appAdS2}. For non-vanishing $\tilde{\phi}$ we find a similar result for $\phi_A>0$, namely 
\begin{equation}\label{FA}
-\beta_T F_{\text{AdS}_2} =  2\pi \left( \phi_0 - \frac{\tilde{\phi}}{2} \right) + \frac{1}{2}\left(2 \phi_b + \tilde{\phi} \right)  \sqrt{\beta_T^2+4\pi^2} ~.
% \beta_A \left( \phi_b+ \frac{\tilde{\phi}}{2} \right)  \phi_b~.
%\frac{\tilde{\phi}  \left(\phi_A - 2 \phi_b - \phi_0\right)  - 2 \left(\phi_0 \, \phi_A + \phi_b^2\right)}{2 \sqrt{\left( \phi_b - \phi_A\right) (\phi_A + \phi_b+ \tilde{\phi} )}}~,
\end{equation}
The specific heat follows readily
\begin{equation}\label{CA}
	C_{\text{AdS}_2} = \left(  \phi_b + \frac{\tilde{\phi}}{2}\right) \, \frac{4 \pi^2 \beta_T^2 }{\left(\beta_T^2+4 \pi ^2\right)^{3/2}}~,
	%\frac{\beta_T (2 \phi_A + \tilde{\phi} )^2 \sqrt{(\phi_b - \phi_A) (\phi_A + \phi_b + \tilde{\phi} )} }{(2 \phi_b + \tilde{\phi} )^2} ~. 
	%\frac{2 \pi  (\text{rb}-\text{ra}) (2 \text{ra}+\text{$\phi $t}) (\text{ra}+\text{rb}+\text{$\phi $t})}{(2 \text{rb}+\text{$\phi $t})^2}
\end{equation}
which is positive  for all $\phi_b > 0$. \\
\\
The other solution has $\phi_D < 0$ such that the region of the geometry near and including the horizon has positive curvature. For this interpolating solution, we have 
%\begin{equation}
%F_{\text{interp}} = \frac{(\tilde{\phi}  + 2 \phi_0 ) \phi_D - 2 \phi_b  (\phi _b + \tilde{\phi})   - \phi_0 \, \tilde{\phi} }{2 \sqrt{\phi_b^2 + \tilde{\phi} \, \phi_b + (\phi_D- \tilde{\phi}) \phi_D}} ~. % = \phi_b  \frac{}{} \sqrt{\beta_T^2 - 4\pi^2}
%\end{equation}
\begin{equation} \label{Finterp1}
-\beta_T F_{\text{interp}} = 2\pi \left( \phi_0 + \frac{\tilde{\phi}}{2} \right)  + \left(\beta_T^2 - 4 \pi^2\right) \sqrt{\frac{(\phi_b + \chi_+)(\phi_b - \chi_- )}{\beta_T^2 - 4 \pi^2}} ~,
%2\pi \left( \phi_0 - \frac{\tilde{\phi}}{2} \right)  +  \phi_b  \left( {2\phi_b + \tilde{\phi}} \right) \sqrt{\frac{\beta_T^2 - 4\pi^2}{4\phi_b^2 + 4\phi_b \tilde{\phi} - \tilde{\phi}^2}}~.
\end{equation} 
where for convenience we have defined $\chi_\pm \equiv \left(\tfrac{\sqrt{2} \pm 1}{2} \right)\tilde{\phi}$. In order to ensure the above expression is real we must lie in one of two regimes. The first, which connects to parameterically large $\phi_b$, is $\phi_b \ge \chi_-$ and $\tfrac{4\pi}{\tilde{\phi}} \sqrt{\phi_b( \phi_b  + \tilde{\phi})} \ge \beta_T \ge 2\pi$, where the upper bound for $\beta_T$ ensures the negativity of $\phi_D$. Given $F_{\text{interp}}$, we can compute the specific heat of the interpolating saddle for this range:
\begin{equation} \label{CInterpunstable}
      C_{\text{interp}} = - \sqrt{(\phi_b + \chi_+)(\phi_b - \chi_-  )} \times  \frac{4 \pi^2 \beta_T^2 }{( \beta_T^2 - 4 \pi^2)^{3/2}} ~. 
      %\frac{\beta_T^D (\tilde{\phi} - 2 \phi_D )^2 \sqrt{\phi_b^2 + \tilde{\phi} \,  \phi_b + (\phi_D - \tilde{\phi} ) \phi_D} }{\tilde{\phi}^2-4 \tilde{\phi} \,  \phi_b - 4 \phi_b^2  } ~.
      %\frac{\beta_T (2 \tilde{\phi}+ \phi_h)^2\sqrt{\phi_b^2 - \phi_h^2}}{(2 \tilde{\phi} + \phi_h)^2 + \phi_h^2- \phi_b^2  } ~.
\end{equation}
%\begin{equation}
      %C_{\text{interp}} =  - \phi_b  \frac{2\phi_b + \tilde{\phi}}{\sqrt{|4\phi_b^2 + 4\phi_b \tilde{\phi} - \tilde{\phi}^2 |}}\frac{4 \pi ^2 \beta_T^2}{ | \beta_T^2-4 \pi ^2 |^{3/2}}~.
      %\frac{\beta_T (2 \tilde{\phi}+ \phi_h)^2\sqrt{\phi_b^2 - \phi_h^2}}{(2 \tilde{\phi} + \phi_h)^2 + \phi_h^2- \phi_b^2  } ~.
%\end{equation}
For this case the specific heat $C_{\text{interp}}$ is negative and the only thermodynamically stable saddle is the Euclidean AdS$_2$ black hole. We thus retrieve, within this range of boundary conditions a near-Nariai black hole horizon with negative specific heat. The second regime is $0< \phi_b < \chi_-$ and $\tfrac{4\pi}{\tilde{\phi}} \sqrt{\phi_b( \phi_b  + \tilde{\phi})} < \beta_T < 2\pi$. For this case the specific heat $C_{\text{interp}}$ is 
\begin{equation} \label{CInterpstable}
      C_{\text{interp}} =  \sqrt{(\phi_b + \chi_+)(-\phi_b +\chi_- )} \times  \frac{4 \pi^2 \beta_T^2}{(4 \pi^2 - \beta_T^2 )^{3/2}} ~,
\end{equation}
which is positive. Note that taking $\phi_b = 0$ in (\ref{CInterpstable}) recovers the result (\ref{Cnariai}) for the black hole in the Nariai geometry up to an overall constant, as expected. Upon solving $\beta_T^D = \beta_T^A = \beta_T$, we see that 
\begin{equation}
	\phi_A = \frac{1}{2} \left(\frac{(2 \phi_b + \tilde{\phi} ) (\tilde{\phi} - 2 \phi_D)}{\sqrt{(2 \phi_b + \tilde{\phi})^2+ 8 \phi_D^2 - 8 \tilde{\phi}  \phi_D}} - \tilde{\phi} \right) , \qquad \phi_b > \chi_- ~.
\end{equation}
A solution for $\phi_A$ for a given $\phi_D$ can only be found for large enough $\phi_b$. 
%\newline\newline
Therefore, in the range $0 < \phi_b < \chi_-$ the interpolating geometry is the only saddle, and is thermodynamically stable. We now explore the case $\tilde{\phi}<0$. 

\subsubsection*{Case 2: $\tilde{\phi} < 0$}

We now consider the thermodynamic properties for $\tilde{\phi} < 0$. The Euclidean AdS$_2$ saddle will be thermodynamically stable, with the same free energy and specific heat as for $\tilde{\phi}>0$, namely (\ref{FA}) and (\ref{CA}) respectively.\\
\\
The interpolating saddle will now permit some additional properties. The free energy will still take the form (\ref{Finterp1}), with the additional restriction coming from the requirement that the metric (\ref{InterpCentaursaddleNeg}) remain positive for all $r$, namely that $\phi_D \leq \chi_+ $. This additional restriction on the value of the dilaton at the horizon modifies the reality conditions for (\ref{Finterp1}). We again find two possible regimes, the first being $\sqrt{2}\pi \left|\tfrac{2 \phi_b + \tilde{\phi}}{\tilde{\phi}}\right| \ge \beta_T \ge 2 \pi$ and $\phi_b \ge - \chi_+$. In this range the interpolating saddle has the same specific heat as in (\ref{CInterpunstable}) and hence is unstable. \\
\\
The other possibility is the regime $\sqrt{2}\pi \left|\tfrac{2 \phi_b + \tilde{\phi}}{\tilde{\phi}}\right| < \beta_T < 2 \pi$ and $0< \phi_b < - \chi_+$. In this case, the heat capacity is given by (\ref{CInterpstable}) and so is thermodynamically stable. Solving $\beta_T^D = \beta_T^A = \beta_T$, we find  
\begin{equation} \label{phiA}
	\phi_A =  \frac{1}{2} \left(  \frac{(2 \phi_b + \tilde{\phi} ) (\tilde{\phi} - 2 \phi_D)}{\sqrt{(2 \phi_b + \tilde{\phi})^2+ 8 \phi_D^2 - 8 \tilde{\phi}  \phi_D}} - \tilde{\phi}  \right) ~.
\end{equation}
Therefore, in this case for a given $\phi_D$, there is always a $\phi_A$ provided that $|\tilde{\phi}|/2 < \phi_b < - \chi_+$. The difference in free energies between the stable interpolating saddle and the AdS$_2$ saddle is given by 
\begin{multline} \label{FdiffInterp}
F_{\text{interp}} - F_{\text{AdS}_2} 
= \frac{ \sqrt{(\phi_b + \chi_+ )(-\phi_b + \chi_-) ( 4 \pi^2 - \beta_T^2  )}}{ \beta_T} 
	+ \frac{(2 \phi_b+ \tilde{\phi} ) \sqrt{\beta_T^2 + 4 \pi^2} - 4 \pi \tilde{\phi} }{2 \beta_T} ~.
\end{multline}
For $\tilde{\phi} < 0$ this expression is positive. To give a numerical example, we can take $\tilde{\phi}= -1$, $\phi_b =1$, and $\beta_T = 5$ which satisfy the conditions on $\beta_T$ and $\phi_b$. These values correspond to taking $\phi_D = - \tfrac{1}{2} - \tfrac{\pi}{\sqrt{ 4 \pi^2 - 25}} \approx -1.33$ in (\ref{phiA}) such  that $\phi_A = \tfrac{1}{2}+ \tfrac{\pi}{\sqrt{ 4 \pi^2 + 25}} \approx 0.89$ lies in the allowed range for $\phi_A$. They lead to a positive difference (\ref{FdiffInterp}).  So, the AdS$_2$ saddle will be thermodynamically favoured over the interpolating saddle. If instead $0< \phi_b < |\tilde{\phi}|/2$, then there will be no AdS$_2$ saddle since $V(\phi_A)< 0$ in this range and we will have the same situation as in (\ref{CInterpstable}), with the interpolating saddle being the only stable solution. 
%For fixed $\phi_b$, we can solve $\beta_T^A = \beta_T^D$ to find $\phi_D$ in terms of $\phi_A$. Solutions only exist for $\phi_b >  \left( \frac{\sqrt{2} - 1}{2} \right) \tilde{\phi}$. At equal temperature, the difference in free energies is 
%\begin{equation}
%	\Delta F = F_{\text{interp}} - F_{\text{AdS}_2} = \frac{(2 \phi_b + \tilde{\phi} ) \sqrt{4 \tilde{\phi}  \phi_b + 4 \phi_b^2 - 8 \tilde{\phi}  \phi_D + 8 \phi_D^2 + \tilde{\phi}^2} + 4 \tilde{\phi}  \left(\phi_D - \phi_b\right) - 4 \phi_b^2 - \tilde{\phi}^2}{4 \sqrt{\phi_b^2 + \tilde{\phi} \,  \phi_b + (\phi_D - \tilde{\phi} ) \phi_D}} ~.
	%-\frac{\phi_h | 2 \phi_h + \tilde{\phi}|}{\sqrt{\phi_b^2 - \phi_h^2}} > 0 ~.
%\end{equation}
%However, only the Euclidean AdS$_2$ solution is globally stable since $F_{\text{AdS}_2}  < F_{\text{interp}}$.
%  However, another choice could be $\phi_b = 0.5$. In this case, the interpolating saddle is the only stable solution.

\subsection{Interpolating the stretched dS$_2$ horizon}

\begin{figure}[H]
 	\centering
 	\includegraphics[width=16cm]{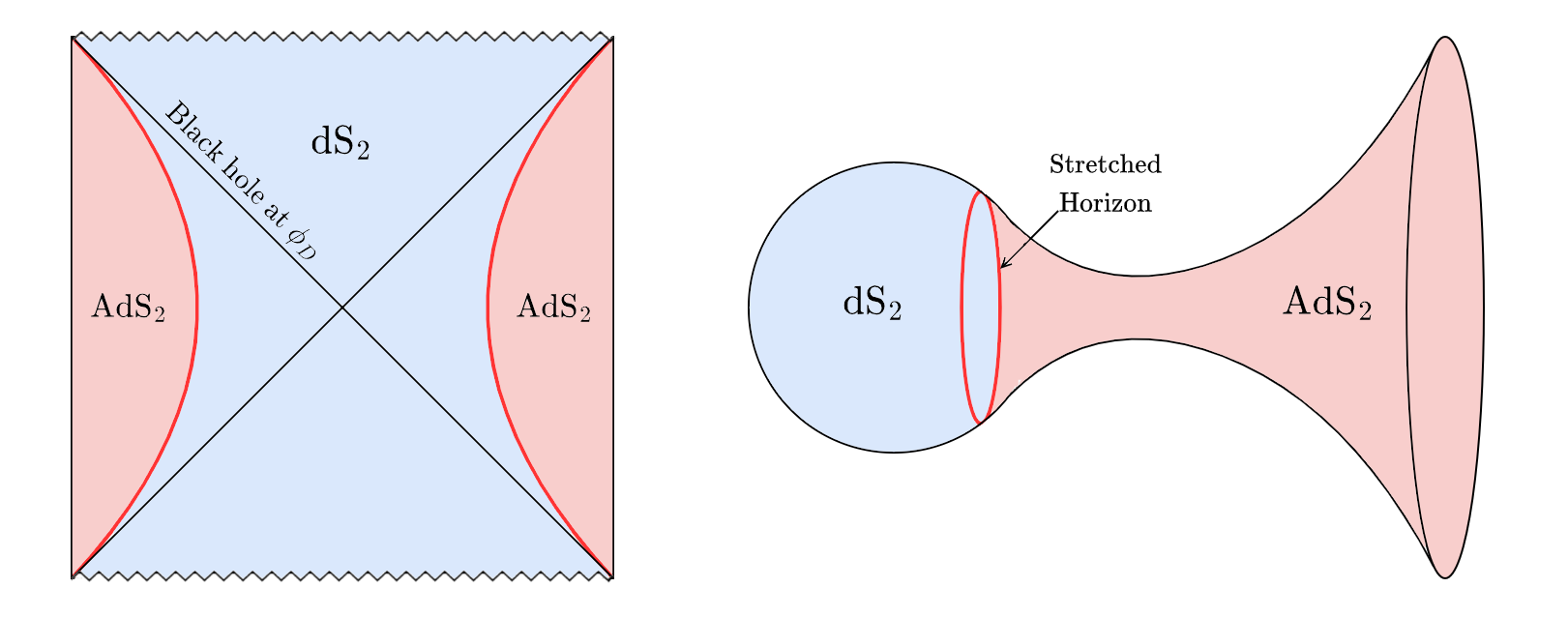}
 	\caption{Left: The Penrose diagram of the interpolating geometry that contains the black hole in a de Sitter static patch in the deep interior and interpolates to an AdS$_2$ boundary. Right: The same geometry in Euclidean signature, where the Euclidean black hole horizon now  caps off at the pole of the two-sphere. }
 	\label{fig:CentaurPenroseDiagram}
 \end{figure}

We end this section by considering the interpolating solution for $\tilde{\phi}< 0$ in some more detail.  For the purpose of our discussion, it will prove useful to  slightly generalise the dilaton potential (\ref{centaurV}) to the following
\begin{equation}
V_\delta(\phi) = 2 \phi  \left( - \Theta(-\phi) + \frac{1}{\delta}    \, \Theta(\phi)   \right) + \tilde{\phi}~,
\end{equation} 
with $\tilde{\phi}<0$, and $\delta$ a small positive number. In order to have an interpolating solution with an asymptotic AdS$_2$ region, the horizon $r_h$ must lie below the critical value $r_\delta \equiv \tilde{\phi} (1 + \sqrt{1+\delta} \, )/2$. At precisely $r_h = r_\delta$ the geometry caps off for a second time at $r = -\tilde{\phi}\, \delta/2$, creating a closed Euclidean universe.
\newline\newline
%For  $\tilde{\phi}< 0$, the dilaton potential (\ref{centaurV}) becomes negative for $\phi \in (\tilde{\phi}/2,-\tilde{\phi}/2)$. Taking $\phi_b > |\tilde{\phi}|/2$, the interpolating saddle will include more than half of the two-sphere before entering the regime of negative curvature.\footnote{For $\phi_b = 0$ and $r_h = \tilde{\phi}$ the geometry is the full two sphere. Here we will require a region of negative curvature in the interpolating geometry so we must push $\phi_b$ further out.} 
If we now tune $r_h$ to be parameterically close to and below $r_\delta$ and take $\phi_b \gg 1$, we find an asymptotically near-AdS$_2$ geometry which includes a significant portion of the two-sphere. The interpolating saddle stemming from $V_\delta(\phi)$ admits a portion of the two-sphere for which the excised region is a disk of area $\sim \delta$. The excised region that would have contained the cosmological dS$_2$ horizon is replaced by a region of negative curvature that reaches all the way out to the AdS$_2$ boundary. The geometry and corresponding Penrose diagram is shown in Figure \ref{fig:CentaurPenroseDiagram}. 
$\,$\newline
An excision of this type is often considered when placing a stretched surface \cite{Susskind:1993if} a small region away from the dS$_2$ horizon and has been explored as a potential holographic screen of de Sitter  in various works including \cite{Banks:2006rx,Bousso:1999cb,Shaghoulian:2022fop,Anninos:2021ihe,Dong:2018cuv}. Though physically appealing, one of the challenges to make the stretched horizon picture precise is that the surface lies in the midst of a gravitating spacetime and it is difficult to obtain concrete observable quantities analogous to those at the boundary of AdS or the flat space $S$-matrix.\footnote{One approach is to set up a type of timelike Dirichlet boundary near the cosmological horizon, as explored in \cite{Hayward:1990zm,Wang:2001gt,Anninos:2011zn,Banihashemi:2022jys,Banihashemi:2022htw}. However, at least in four and higher spacetime dimensions, caution must be exercised since this leads to various instabilities and issues with well-posedness of the Dirichlet problem \cite{Anderson:2007jpe,An:2021fcq}.}  Instead of placing a holographic theory at the stretched horizon, we might then view the interpolating geometry as an ultraviolet completion of the stretched dS$_2$ horizon with a near-AdS$_2$ boundary. %residing near $r_\delta $
\newline\newline
As it stands, the interpolating geometries we have discussed so far have negative specific heat whenever they are endowed with an asymptotically AdS$_2$ boundary. This can be ameliorated by bringing the near-AdS$_2$ boundary into the interior, by introducing a Dirichlet wall \cite{Gross:2019ach} along the lines we have discussed in the previous sections. This comes at the cost of sharp AdS/CFT type observables. In the next section we discuss a simple generalisation that admits a stretched dS$_2$ horizon that is capped off by a horizon with positive specific heat in the deep interior while preserving the asymptotic near-AdS$_2$ boundary. 
% Given this setup, one can exploit AdS/CFT type observables at the near AdS$_2$ boundary to define the interior of the interpolating geometry. We leave a detailed analysis of such observables to future work. 

%\subsection{Free energy}
%For a given temperature, there can be multiple possible black hole solutions, as shown in figure \ref{fig:CentaurPotential}, where for 
%\begin{equation}
%    T = \frac{V(\phi_1)}{4\pi} = \frac{V(\phi_2)}{4\pi} ~,
%\end{equation}
%there are two candidate black holes at $\phi_1$ and $\phi_2$. The free energy is 
%\begin{equation}
%	F(\phi_h) = \frac{-\left(\phi _h+\phi _0\right) | \phi _h- 2 \tilde{\phi} | -\phi _b^2+\phi _h^2}{\sqrt{\phi_b^2 - \phi_h^2}} ~.
%\end{equation}
\section{Double interpolating geometries}\label{disec}
In this section we will propose a geometry that has some of the benefits of the  interpolating geometry discussed in the previous section while also having the virtue of a positive specific heat, ensuring that the solution is locally thermodynamically stable. For the sake of concreteness, we will consider a theory with the following dilaton potential
\begin{equation} \label{SandPot}
    V(\phi) = \begin{cases}
  2 \phi + \tilde{\phi} - 4 x~, \quad\quad & \phi \leq x~, \\
 2 | \phi | + \tilde{\phi}~, \quad\quad & x< \phi ~, \\
\end{cases}
\end{equation}
where $x < 0$. As in the previous section, the potential can be viewed as an idealisation of a smooth potential where we have only retained the piecewise linear pieces. 
\newline\newline
We now discuss the asymptotically AdS$_2$ Euclidean saddles and their thermodynamic properties, while relegating the discussion with finite boundary to appendix \ref{Double Interpolating Geometry finite}. Here, we keep the slopes of each linear piece of $V(\phi)$ to have the same magnitude for the sake of simplicity. More generally, they can be taken to be different. 

\subsection{Geometry}

\begin{figure}[H]
	\centering
	\includegraphics[width=10cm]{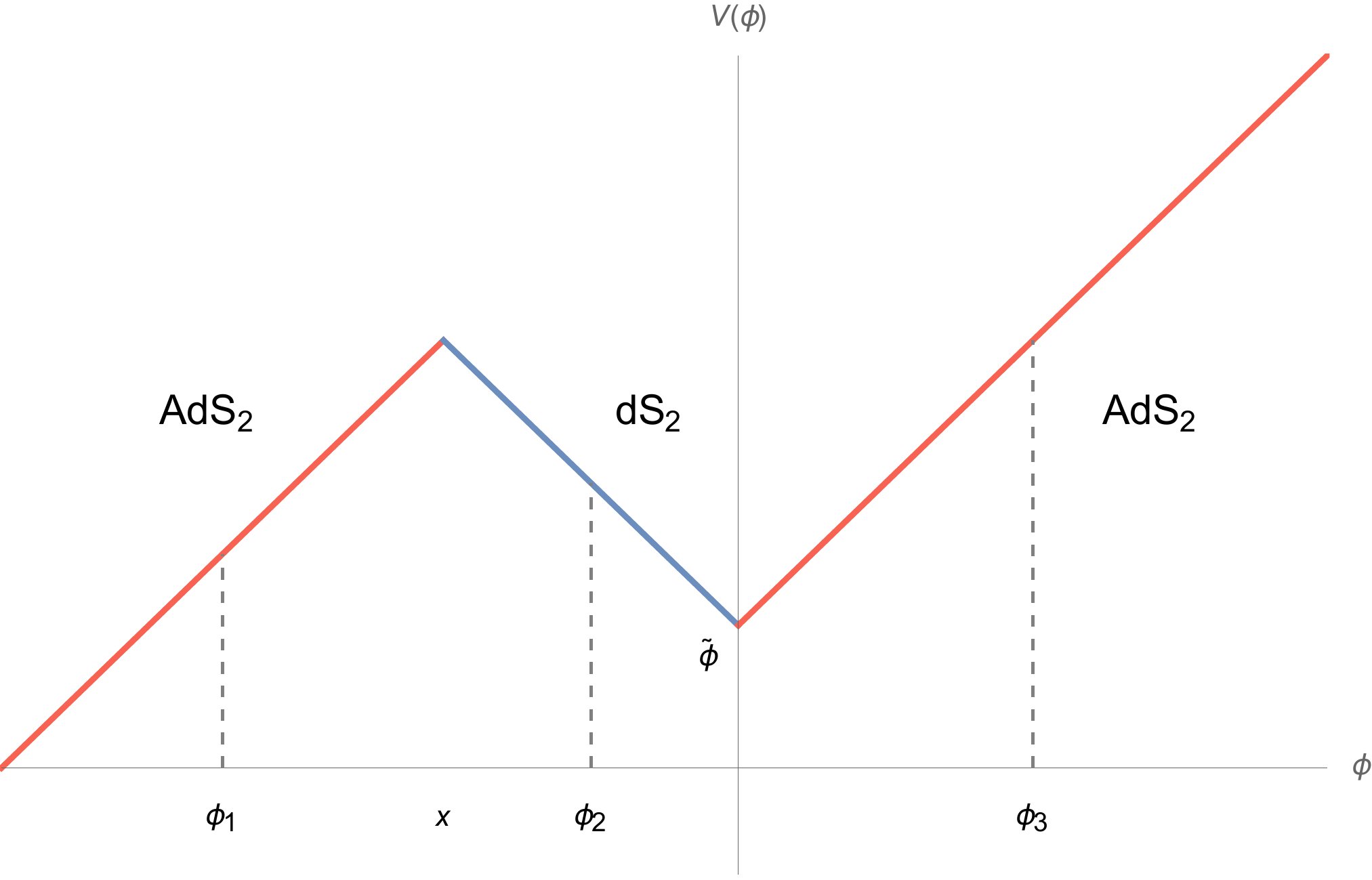}
	\caption{Plot showing the dilaton potential (\ref{SandPot}) with $\tilde{\phi} > 0$. This potential gives an AdS$_2$ geometry for $\phi \leq x$ and $\phi \geq 0$ but a dS$_2$ static patch in the region $x \leq \phi \leq 0$. At a given temperature, there may be up to three candidate black hole solutions. }
	\label{fig:SandwichPotential}
\end{figure}

The geometry is given by merging the interpolating geometry (\ref{InterpCentaursaddle}) to a second AdS$_2$ region in the deep interior at the distance $r = x$ as is shown in Figure \ref{fig:SandwichPotential}. In the range  $\tilde{\phi} < V(\phi) < \tilde{\phi} - 2x$, we therefore have three possible saddles which we label $\phi_1 < \phi_2 <0 < \phi_3$. Outside of this range, the solution will only contain a stable AdS$_2$ black hole saddle if $\tilde{\phi} - 2x <V(\phi)$, or only the stable double interpolating geometry if $V(\phi) < \tilde{\phi}$.
\subsubsection*{Case 1: $\tilde{\phi} \ge 0$}
The first saddle is for $r_1 < x$ and the metric for this system will have coefficients given by 
\begin{equation}  \label{Doubleintepsaddle1}
	N(r, r_1) = 
	\begin{cases} 
		(r - r_1)(r + r_1 + \tilde{\phi} - 4 x)~, &  \tfrac{4x - \tilde{\phi}}{2} \le r_1 \leq r \leq x ~,\\
		-r^2 + \tilde{\phi} \,r - 2 x^2 + r_1 ( 4x - r_1 - \tilde{\phi}) ~, & x < r \leq 0 ~,\\
		r^2 + \tilde{\phi} \, r - 2x^2 + r_1 (4x - r_1 - \tilde{\phi})~, & 0 < r ~,
	\end{cases} 
\end{equation}
where the first condition on $r_1$ ensures that $V(r_1) \ge 0$. This describes a geometry which is AdS$_2$ in the deep interior, flowing to a piece of the Euclidean dS$_2$ static patch, and then flowing to another AdS$_2$ region near the boundary. The size of the static patch region is controlled by the two free parameters $x$ and $\tilde{\phi}$.\footnote{By further adjusting the slopes of the linear pieces in $V(\phi)$, the AdS$_2$ in the deep interior can be made parameterically small.}\\
\\
The second saddle has a Euclidean horizon located in the range $x \le r_2 <  0$. This is the interpolating geometry described in Section \ref{(A)dS$_2$ Interpolating geometries} where the metric is precisely (\ref{InterpCentaursaddle}) with $r_D = r_2$. The final saddle has a Euclidean horizon located at $0 \leq r_3 \leq r$. For this range, the metric is (\ref{AdSCentaursaddle}) with $r_A = r_3$ which again is the AdS$_2$ saddle described in the previous section. 
 
\subsubsection*{Case 2: $\tilde{\phi} < 0$}
\begin{figure}[H]
	\centering
	\includegraphics[width=15cm]{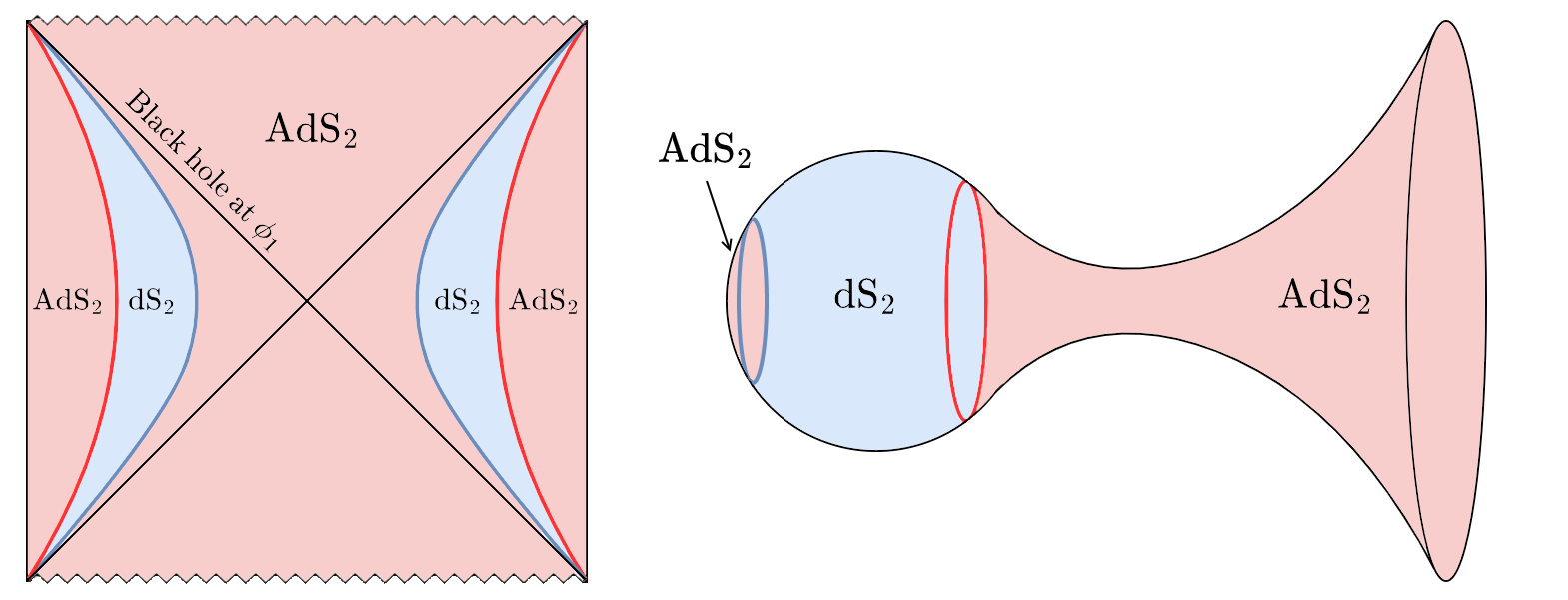}
	\caption{Left: Penrose diagram of the double interpolating geometry. Right: corresponding Euclidean geometry.}
	\label{fig:doubleinterpolating}
\end{figure}

In this case, the geometries are as above, but each saddle has an added restriction. Firstly, in order to have values of $r_h = r_1, r_2$ with $V(r_h)>0$, we require $V(x)>0$ such that $\tilde{\phi} > 2x$. This ensures that the lower bound on $r_1$ in (\ref{Doubleintepsaddle1}) remains less than $x$ when $\tilde{\phi}<0$. To ensure the positivity of the metric (\ref{Doubleintepsaddle1}) we must further have $4 r_1( 4x - \tilde{\phi} - r_1) > 8 x^2 + \tilde{\phi}^2$.  \\
\\
The restriction for the second saddle at $r_2$ is the same as in (\ref{InterpCentaursaddleNeg}), namely $x \leq r_2 \leq \chi_+$ and for the third saddle at $r_3$ we have the same added restriction as in (\ref{AdSCentaursaddleNeg}) that $|\tilde{\phi}|/ 2 \le r_3$.
\newline\newline
As for the case of the interpolating solution with $\tilde{\phi}<0$ discussed in the previous section, we can view the completion of the excised two-sphere as a way to push the stretched dS$_2$ horizon all the way to the near-AdS$_2$ boundary. This is shown in Figure \ref{fig:doubleinterpolating}. As we now explore, in the current setup the thermal stability properties are improved as compared to the previous case. 

\subsection{Thermodynamic properties}

Let us introduce the notation $\phi_1 = r_1$, $\phi_2 = r_2$ and $\phi_3 = r_3$ as the values of the dilaton at each horizon. In what follows we will take $\phi_b \gg 1$ so that the boundary is asymptotically AdS$_2$. The finite boundary case is treated in appendix \ref{Double Interpolating Geometry finite}. In this limit we can employ the asymptotic formulas (\ref{asympESC}). Along with these, in the large $\phi_b$ limit the difference in free energy of any two saddles $\phi_1$ and $\phi_2$ is given by \cite{Witten:2020ert}
\begin{equation} \label{Fdiffasympt}
	\Delta F = F_2 - F_1 = \frac{1}{2  \phi_b} \left[  N(\phi_2, \phi_1) + V(\phi_1) ( \phi_1 - \phi_2 )  \right] ~.
\end{equation}
As $\phi_b \rightarrow  \infty$, we have that $\beta_T  \approx  \beta \phi_b$, and so from the definition (\ref{beta}) we must also have $V(\phi_1) = V(\phi_2)$. 
\subsubsection*{Case 1: $\tilde{\phi} \ge 0$}
For $\phi_b \gg 1$ the first saddle at $\phi_1$ will have heat capacity 
\begin{equation} \label{asymC1}
	C_1 = \pi ( 2 \phi_1 + \tilde{\phi} - 4x ) ~,
\end{equation}
which is positive due to the lower bound on $r_1$ in (\ref{Doubleintepsaddle1}). As promised, this geometry has the benefit of being thermodynamically stable while also possessing an asymptotically AdS$_2$ boundary. The ground state with vanishing temperature is given by the configuration starting at the point $\phi_1 = 2x-\tilde{\phi}/2$ where the dilaton potential crosses the $\phi$-axis. 
\newline\newline
The thermodynamic properties of the saddles at $\phi_2$ and $\phi_3$ were described in Section \ref{Interpolating Thermodynamic properties}. For large $\phi_b$, the saddle at $\phi_2$ will have negative heat capacity (\ref{CInterpunstable}) and hence will be thermodynamically unstable. The saddle at $\phi_3$ will have heat capacity given by (\ref{CA}) and so will be thermodynamically stable. To see which saddle is thermodynamically favoured, we can use (\ref{Fdiffasympt}) with $V(\phi_1) = V(\phi_3)$ leading to 
\begin{equation} \label{asympFdiff}
	\Delta F = F_3 - F_1 = \frac{x}{\phi_b} \left( 2\phi_1 - 3 x \right) ~. 
\end{equation}
This will be positive if $2 \phi_1 < 3x $ and negative otherwise. 
\newline\newline
Thus, for  temperatures satisfying $\beta > {4\pi}/({\tilde{\phi}-x})$, the double interpolating solution containing a large portion described by Euclidean dS$_2$ dominates the thermodynamics in this model. Once $\beta$ reaches the critical value $\beta = {4\pi}/({\tilde{\phi}-x})$ we have a first order phase transition to the Euclidean AdS$_2$ black hole. In this case, the interpolating solution is  metastable.

\subsubsection*{Case 2: $\tilde{\phi} < 0$}

The model has two vanishing temperature configurations, one at $\phi_1 = 2x-\tilde{\phi}/2$, with $x$ restricted to yield a positive metric,  and the other at $\phi_3 = -\tilde{\phi}/2$. At finite temperature,  the specific heat  of the $\phi_1$ saddle is  (\ref{asymC1}) which remains positive. For $\phi_b \gg 1$, the saddle at $\phi_2$ is again unstable and hence we again only have thermally stable saddles at $\phi_1$ and $\phi_3$ with their difference again given by (\ref{asympFdiff}). 
\newline\newline
Given that $\Delta F$ does not depend on $\tilde{\phi}$, the sign of $\tilde{\phi}$ will not effect which saddle dominates. Indeed, the model exhibits a first order phase transition at the critical temperature $\beta= {4\pi}/({\tilde{\phi}-x})$ between a low temperature phase dominated by the double interpolating saddle and a high temperature phase dominated by the Euclidean AdS$_2$ black hole. 
%\section{Flow of three AdS$_2$ spacetimes}
%In this section we will study a dilaton gravity model that flows between three AdS$_2$ geometries with different curvature scales. The dilaton potential for sych a theory is 
%\begin{equation}
%    V(\phi) = \begin{cases}
%    a \, \phi + b &\phi \leq x \\
%    c \, \phi + b + (a - c) x & x \leq \phi \leq \tilde{\phi} \\
%    d \, \phi + b + (a - c) x + (c - d) y & \tilde{\phi} < \phi ,
%    \end{cases}
%\end{equation}
%where $a, c ,d >0$. Then by the equation of motion (\ref{Riccieom}), the Ricci curvature scalar in each part of the geometry will be positive.  
\newline\newline
To summarise, we have established the existence of dilaton-gravity models permitting locally (and even globally) thermally stable asymptotically near-AdS$_2$ geometries that encode a significant portion of the Euclidean dS$_2$ static patch in their interior. Slightly generalising the potential (\ref{SandPot}) to have different slopes in each linear regime, one can have asymptotically near-AdS$_2$ geometries that contain a stretched de Sitter horizon parameterically close to the actual cosmological dS$_2$ horizon. The geometries end at an AdS$_2$ black hole type horizon with positive specific heat. An alternative  way to stabilise the interpolating geometries in the deep interior might be to include an end-of-the-world brane. Such a scenario would also be interesting to explore.

\section{Outlook}\label{outlooksec}

We have explored a variety of dilaton-gravity models whose solution space includes interpolating solutions containing a portion of the static patch of two-dimensional de Sitter space. In judiciously chosen circumstances, the stability properties of these geometries can be locally and even globally stable. In particular, the double interpolating solutions discussed in Section \ref{disec} have asymptotically near-AdS$_2$ solutions which contain a portion of the dS$_2$ static patch. Moreover, these solutions have positive specific heat and can be thermodynamically dominant over the AdS$_2$ black hole.
\newline\newline
Given the presence of a near-AdS$_2$ boundary, one is tempted to investigate whether these models permit a microphysical realisation in terms of a AdS$_2$/CFT$_1$ type picture, or some other ultraviolet completion of dilaton-gravity. One such approach might be to consider the more general relation \cite{Saad:2019lba,Witten:2020wvy,Maxfield:2020ale} between dilaton-gravity models with general dilaton potential and matrix models. The challenge here is that the type of dilaton-potentials discussed in those works take a somewhat specific form
\begin{equation}\label{deformedV}
V_f(\phi) = 2\phi + \int_\pi^{2\pi} d\alpha f(\alpha) e^{-\alpha\phi}~,
\end{equation}
with $f(\alpha)$ small.  Although it is suggested that the class of perturbative deformations in (\ref{deformedV}) might span to the larger class of holomorphic functions, $V_f(\phi)$ is subject to non-perturbative corrections. Whenever $x$ is small and $\tilde{\phi}$ is positive, the potential (\ref{SandPot}), or a smoothened out version, can be viewed as a perturbation $\delta V(\phi)$ of $V(\phi)= 2\phi$. So, provided that the deformed potentials (\ref{deformedV}) span a sufficiently large space of deformations, one might identify a corresponding matrix model. Employing the results of \cite{Witten:2020wvy,Maxfield:2020ale}, the corresponding eigenvalue distribution to linear order in the deformation (upon shifting $\phi$ such that $V_f(0)=0$) reads
\begin{equation}\label{rhocompact}
\rho(\lambda) = e^{S_0}\left( \frac{\sinh 2\pi\sqrt{\lambda}}{4\pi^2} + \frac{e^{2\pi\sqrt{\lambda}}\delta V(\sqrt{\lambda}) +e^{-2\pi\sqrt{\lambda}}\delta V(-\sqrt{\lambda})}{8\pi\sqrt{\lambda}} + \ldots   \right)~.
\end{equation}
For the models we consider $\rho(\lambda)$ increases indefinitely or at least up to some large cutoff, reflecting the near-AdS$_2$ boundary. For a closed Euclidean universe, as suggested by (\ref{rhocompact}), perhaps one should take $\rho(\lambda)$ to fall back to a vanishing value \cite{Anninos:2021ydw,Anninos:2020geh,Anninos:2022ujl}, reflecting the presence of two horizons in dS$_2$ \cite{Anninos:2017hhn,Dong:2018cuv}. It would be interesting to explore such matrix models further. 
\newline\newline
More generally, one could ask whether the SYK model \cite{Sachdev:1992fk,Kitaev:2017awl,Maldacena:2016hyu} permits deformations leading to theories such as (\ref{SandPot}). Since the vacuum is no longer pure AdS$_2$ in the interior, we expect that the SYK model is deformed by some relevant deformation. Relevant deformations are indeed permitted in SYK, and were explored in \cite{Anninos:2020cwo} where flows between two near-CFTs were identified. In particular, this was shown for the sum of two SYK Hamiltonians, $\hat{H}_{\text{tot}} = \hat{H}_q+ s \, \hat{H}_{q/2}$, where 
\begin{equation}
\hat{H}_q = {i^{q/2}} \sum_{i_1\leq \ldots\leq i_q \leq N} J_{i_1,\ldots,i_q} \psi_{i_1} \ldots \psi_{i_q}~,
\end{equation}
with $J_{i_1,\ldots,i_q}$ independently drawn from a suitable Gaussian ensemble, and $\psi_i$ with $i=1,\ldots,N$ being Majorana fermions. Provided $s$ is sufficiently small, $\hat{H}_{\text{tot}}$ flows between two near-CFT regions. 
%\newline\newline
For the theory  (\ref{SandPot}) we have an intermediate region of near-dS$_2$ between the two near-AdS$_2$ regions near the boundary and in the deep interior. It is likely that this will require adding an additional term to $\hat{H}_{\text{tot}}$.  We leave the exploration of such constructions for future work.

\section*{Acknowledgements} 
%\newline\newline
It is a great pleasure to acknowledge Tarek Anous, Dami\'an Galante, Diego Hofman, Beatrix M\"uhlmann, Ben Pethybridge, Sameer Sheorey, Edgar Shaghoulian, and Eva Silverstein for useful discussions. D.A. is funded by the Royal Society under the grant The Atoms of a deSitter Universe. E.H. is funded by an STFC studentship ``Aspects of black hole and cosmological horizons". The authors would also like to thank the participants of the Corfu Summer Institute Workshop ``Features of a Quantum de Sitter Universe". 

%\newpage

\appendix

\section{Thermodynamics of finite boundary geometries}\label{appAdS2}

In this appendix we consider pure JT gravity and the dilaton-gravity model arising from the dimensional reduction of four-dimensional Einstein gravity in the presence of a Dirichlet boundary. 

\subsection{AdS$_2$ JT gravity}
The dilaton potential that gives rise to an AdS$_2$ geometry is $V(\phi) = 2 \phi$. We can see that this potential results in the Euclidean AdS$_2$ black hole metric (which is the Poincar\'e disk) by using equation (\ref{V-Areleation}):
\begin{equation}
    ds^2 = (r^2 - r_h^2 ) d\tau^2 + \frac{dr^2}{(r^2 - r_h^2)}~, \qquad r \in [r_h,\infty) ~,
\end{equation}
where $r_h>0$ and $\tau \sim \tau + \beta$. The on-shell action is 
\begin{equation}
\begin{split}
    \log Z_{\text{AdS}_2} %&=-() \frac{\phi_0 \beta }{2} \int_{r_h}^{r_b} d r A''(r) - \frac{\phi_0 \beta}{2} A'(r_b)  - \frac{\beta}{2} A'(r_b) \phi_b )\\ 
    &= 2 \pi  \phi_0  +  \beta \, \phi_b^2 ~,
\end{split}
\end{equation}
with the Euclidean time periodicity $\beta = 2\pi/r_h$. The Tolman temperature is
\begin{equation}
    \beta_T  = \frac{2\pi}{\phi_h} \sqrt{r_b^2 - \phi_h^2}~,
\end{equation}
where, as in the previous sections, $\phi_h = r_h$. One thus finds \cite{Lemos:1996bq}
\begin{equation}
	 - \beta_T F_{\text{AdS}_2} = 2 \pi \phi_0 + \phi_b \sqrt{4 \pi^2 + \beta_T^2}~ . 
\end{equation}
Using the expressions (\ref{energy}), (\ref{entropy}), and (\ref{heatcapacity}), we find the following thermodynamic quantities
\begin{equation}
    E_{\text{AdS}_2}  = - \sqrt{\phi_b^2 - \phi_h^2}~, \quad\quad S_{\text{AdS}_2}  = 2 \pi \phi_0 + 2 \pi \phi_h~, \quad\quad C_{\text{AdS}_2}  = \frac{4 \pi^2   \phi_b \, \beta_T^2}{\left(\beta_T^2+4 \pi^2\right)^{3/2}}~.
 %   \frac{\phi_h^2 \, \beta_T}{ \phi_b^2} \sqrt{\phi_b^2 - \phi_h^2}~.
\end{equation}
We note that $C_{\text{AdS}_2}$ is manifestly positive since $\phi_b > 0$. If we take $\phi_b \gg 1$, we find the leading order expressions
\begin{equation}
    E_{\infty}  = -\phi_b + \frac{2\pi^2}{\phi_b\beta^2}~, \quad\quad S_{\infty}  = 2 \pi \phi_0 + \frac{4 \pi^2}{\beta}~, \quad\quad C_{\infty}  = \frac{4\pi^2}{\beta}~,
\end{equation}
in agreement with the known expressions (see, for example,  \cite{Witten:2020ert}) for asymptotically near-AdS$_2$ up to a physically inconsequential shift in $E_{\infty}$. We note that in this limit thermodynamical variations are with respect to $\beta$ which is defined entirely at the horizon. 

\subsection{Schwarzschild dilaton gravity}
The dilaton potential that appears in the dimensional reduction of the four-dimensional Schwarzschild solution is \cite{Cavaglia:1998xj}
\begin{equation}
    V(\phi) = \frac{1}{2 \sqrt{\phi}}, \qquad \phi \geq 0 ~. 
\end{equation}
Using (\ref{V-Areleation}) we find the two-dimensional Euclidean metric
\begin{equation}
    ds^2 = (\sqrt{r} - \sqrt{r_h}) \, d\tau^2 + \frac{dr^2}{(\sqrt{r} - \sqrt{r_h})}~. % = \frac{1}{2 } \int_{r_h}^r dr' \sqrt{\frac{1}{ r '}} 
\end{equation}
%The bulk action is 
%\begin{equation}
%\begin{split}
%    S_{\text{bulk}} &=   - \frac{\beta}{2} \int_{r_h}^{r_b} dr \left( -  r A''(r) + \frac{1}{2 \sqrt{r}} \right)  \\
%    &= - \frac{3 \beta}{4}  \left( \sqrt{r_b} - \sqrt{r_h}\right) ~. 
%\end{split}
%\end{equation}
%The boundary term is 
%\begin{equation} 
%    S_{\text{bdy}} = - \frac{\beta}{2} A'(r_b) \phi_b = - \frac{\beta \, \phi_b}{4 \sqrt{r_b}} ~.
%\end{equation}
The on-shell Euclidean action yields
\begin{equation}
\begin{split}
    \log Z_{\text{flat}} &= 2 \pi \phi_0 + \beta \left( \sqrt{\phi_b} -\frac{3}{4} \sqrt{\phi_h}  \right) ~, 
\end{split}
\end{equation}
where $\beta = 8 \pi \sqrt{r_h}$ is the periodicity in Euclidean time and $\phi_h = r_h$. The Tolman temperature is 
\begin{equation}
    \beta_T  =  8 \pi \sqrt{\phi_h} \sqrt{  \sqrt{r_b} - \sqrt{\phi_h}} ~.
\end{equation}
The thermodynamic properties of Schwarzschild in $2$d with a finite boundary are found to be
\begin{equation}\label{SchwHC}
    E_{\text{flat}} = - \sqrt{  \sqrt{\phi_b} - \sqrt{\phi_h}}~, \quad S_{\text{flat}} = 2 \pi \left( \phi_0 +  \phi_h \right)~, \quad C_{\text{flat}} = \frac{8 \pi \phi_h ( \sqrt{\phi_b} - \sqrt{\phi_h})}{3 \sqrt{\phi_h} - 2 \sqrt{\phi_b}}~.
\end{equation}
Here we see further evidence of a finite boundary resulting in a transition from positive to negative heat capacity. Since $\phi_b > \phi_h >0$, the numerator of $C_{\text{flat}}$ in (\ref{SchwHC}) is always positive, but the sign of the denominator depends on the location of the wall with respect to $\phi_h$. This is the phenomenon observed by York for Schwarzschild black holes in a Dirichlet box \cite{Hawking:1982dh, York:1986it}. 

\section{Double Interpolating Geometry with finite boundary} \label{Double Interpolating Geometry finite}

In this appendix, we consider the Dirichlet problem,  with  $\beta_T$ and $\phi_b$ fixed, for the dilaton-gravity theory with potential (\ref{SandPot}).
\newline\newline 
For $\phi_b>0$, the temperatures of each saddle are 
\begin{eqnarray}
	\beta_T^1 &=& \frac{4\pi}{|2 \phi_1 + \tilde{\phi} - 4x|} \sqrt{r_b^2 + \tilde{\phi} \, r_b - 2x^2 + \phi_1 (4x - \phi_1 - \tilde{\phi})} ~, \\
	\beta_T^2 &=& \frac{4\pi}{|\tilde{\phi} - 2 \phi_2|} \sqrt{r_b^2 + \phi_2^2 + \tilde{\phi}(r_b - \phi_2)} ~, \\
	\beta_T^3 &=& \frac{4\pi}{|\tilde{\phi} + 2 \phi_3|} \sqrt{r_b^2 - \phi_3^2 + \tilde{\phi}(r_b - \phi_3)} ~.
\end{eqnarray}
\subsubsection*{Case 1: $\tilde{\phi} \ge 0$}
We require $\phi_b > 0$ to have a geometry containing the full, double interpolating solution (\ref{Doubleintepsaddle1}). The free energy of the first saddle at $\phi_1$ is 
\begin{equation} \label{F1}
	-\beta_T F_{1} =  2 \pi \left( \phi_0 + 2 x  - \frac{\tilde{\phi}}{2} \right) + \frac{1}{2} \sqrt{(\beta_T^2 + 4 \pi^2 )\left((2 \phi_b + \tilde{\phi} )^2 + 8 x^2 - 8 x \tilde{\phi}\right) }  ~.
\end{equation}
The properties of the second and third saddles have been described in section \ref{Interpolating Thermodynamic properties}, with their free energies given by (\ref{Finterp1}) and (\ref{FA}) respectively. From equation (\ref{heatcapacity}), the heat capacity of the double interpolating saddle is
\begin{equation} \label{C1}
	C_1 = \frac{2 \pi^2 \beta _T^2 \sqrt{(2 \phi_b + \tilde{\phi})^2+8 x^2 -8 x \tilde{\phi} }}{(\beta_T^2 + 4 \pi^2)^{3/2}}> 0~,
\end{equation}
and hence the saddle is stable, similarly to the case with an asymptotic AdS$_2$ boundary. The heat capacity of the second saddle $C_2$ will be either (\ref{CInterpunstable}) or (\ref{CInterpstable}) depending on the ranges of $\beta_T$ and $\phi_b$ as described in section \ref{Interpolating Thermodynamic properties}. The third saddle $C_3$ will have specific heat (\ref{CA}). \\
\\
First take the case $ \tfrac{4\pi}{\tilde{\phi}} \sqrt{\phi_b( \phi_b  + \tilde{\phi})} \ge \beta_T \ge 2\pi$ and $\phi_b \ge \chi_-$ such that the second saddle has heat capacity (\ref{CInterpunstable}) and is unstable. Then comparing the free energies of the first and third saddle we find
\begin{multline} \label{Fdiff31}
	\Delta F = F_{3} - F_{1}\\
	= \frac{1}{2\beta_T} \left(8 \pi x - \sqrt{\beta_T^2 + 4 \pi^2} \left(2 \phi_b + \tilde{\phi}\right)+  \sqrt{\left(\beta_T^2+ 4 \pi^2\right) \left((2 \phi_b + \tilde{\phi} )^2+8 x^2 - 8 x \tilde{\phi} \right)}\right) ~.
\end{multline}
This can either be positive or negative depending on the range of $x$. The difference is positive if 
\begin{equation} \label{xcond}
	x < \frac{1}{\beta_T^2 - 4 \pi^2} \left( (\beta_T^2 +4 \pi^2)\tilde{\phi} - 2 \pi \sqrt{(\beta_T^2 + 4 \pi^2)(4 \phi_b^2 + 4 \phi_b \tilde{\phi} + \tilde{\phi}^2)} \right)~, 
\end{equation}
and negative otherwise. For example, take $\tilde{\phi}= 1, \, \phi_b = 1$ and $\beta_T = 4\pi$ to satisfy the conditions on $\phi_b$ and $\beta_T$. Then if $x< \tfrac{5}{3} - \sqrt{5}$ then (\ref{Fdiff31}) will be positive and otherwise it will be negative. Thus, for certain temperatures, the double interpolating saddle will be dominant.\\
\\
When instead we take $ \tfrac{4\pi}{\tilde{\phi}} \sqrt{\phi_b( \phi_b  + \tilde{\phi})} < \beta_T < 2\pi$ and $\phi_b < \chi_-$, we have a stable saddle at $\phi_2$ and no longer have a saddle at $\phi_3$. In this case the difference in free energies is 
\begin{multline}  \label{Fdiff21}
	\Delta F = F_{2} - F_{1} = \frac{1}{2 \beta_T} \left(4\pi (2x - \tilde{\phi}) + \right. \\ \left.
	  \sqrt{(\beta_T^2 + 4 \pi^2 )(8 x^2 - 8 x \tilde{\phi} + (2 \phi_b + \tilde{\phi})^2)} + \sqrt{(4\pi^2 - \beta_T^2)(\tilde{\phi}^2 - 4 \phi_b \tilde{\phi} - 4 \phi_b^2)}\right) ~,
\end{multline}
which is negative for this range of $\beta_T$. Therefore, the interpolating saddle will dominate over the double interpolating one.%As an example, take $\tilde{\phi}= 20,\, \phi_b = 1, \,\beta_T = \pi$ and $x = -1$. T

\subsubsection*{Case 2: $\tilde{\phi} < 0$}
In this case, the free energy and heat capacity will still be given by (\ref{F1}) and (\ref{C1}) respectively. To ensure that these expressions are real, we must have $ (2 \phi_b + \tilde{\phi})^2 > 8x ( \tilde{\phi}- x)$. As in the case $\tilde{\phi} < 0$ in section \ref{Interpolating Thermodynamic properties}, the change in sign of $\tilde{\phi}$ restricts the possible ranges of $\beta_T$. Again, the first case is when $\sqrt{2}\pi \left|\tfrac{2 \phi_b + \tilde{\phi}}{\tilde{\phi}}\right| \ge \beta_T \ge 2 \pi$ and $\phi_b \ge - \chi_+$ where the $\phi_2$ saddle is again unstable.  The difference between the free energies of the third and first saddle is (\ref{Fdiff31}). For $\tilde{\phi} <0$ this can again be either positive if the condition (\ref{xcond}) is satisfied, or negative otherwise.\\
\\
If instead, $\sqrt{2}\pi \left|\tfrac{2 \phi_b + \tilde{\phi}}{\tilde{\phi}}\right| < \beta_T < 2 \pi$ and $|\tilde{\phi}|/2< \phi_b < - \chi_+$, then all three saddles at $\phi_1$, $\phi_2$ and $\phi_3$ are stable. As seen in equation (\ref{FdiffInterp}), in this case the AdS$_2$ saddle at $\phi_3$ is thermodynamically favoured over the interpolating one at $\phi_2$. The difference in the free energies between $\phi_3$ and $\phi_1$ is again given by (\ref{Fdiff31}), which in this case will be negative, and so the AdS$_2$ saddle at $\phi_3$ is favoured. However, if $0 < \phi_b < |\tilde{\phi}|/2$ then there is no $\phi_3$ saddle and the  difference between the $\phi_1$ and $\phi_2$ saddles is given by (\ref{Fdiff21}), which again is negative under the asumptions we have here. Therefore, the interpolating saddle will again dominate over the double interpolating one.
\bibliography{BibliographyJT}{}
\bibliographystyle{unsrturl}
\end{document}